\documentclass[showpacs,amsmath,amssymb,aps,twocolumn,superscriptaddress,prx, footinbib]{revtex4-2}

\usepackage{mathtools}
\usepackage{algorithm}
\usepackage{amsthm}
\usepackage{algpseudocode}
\usepackage{grffile} 

\usepackage{ulem} 
\usepackage{comment}

\usepackage{float}
\usepackage{dcolumn}
\usepackage{bm}
\usepackage[colorlinks=true]{hyperref}

\hypersetup{
	colorlinks=true,
	linkcolor=blue,
	urlcolor=cyan,
}

\usepackage{color}
\definecolor{ForestGreen}{RGB}{34, 139, 34}

\newcommand{\slr}{$\mathrm{SL}(2,\mathbb{R})$}
\newcommand{\slc}{$\mathrm{SL}(2,\mathbb{C})$}

\newcommand{\affA}{State Key Laboratory of Artificial Microstructure and Mesoscopic Physics, School of Physics, Peking University, Beijing 100871, China}
\newcommand{\affB}{Max Planck Institute for the Physics of Complex Systems, N\"othnitzer Str.~38, 01187 Dresden, Germany.}
\newcommand{\affC}{Department of Physics, Princeton University, Princeton, New Jersey 08544, USA}

\begin{document}
		
\title{Complex and tunable heating in conformal
field theories with structured drives via classical
ergodicity breaking}

\author{Liang-Hong Mo}
 \affiliation{\affB}
  \affiliation{\affA}
     \affiliation{\affC}

\author{Roderich Moessner}
 \affiliation{\affB}

 	\author{Hongzheng Zhao}
\email{hzhao@pku.edu.cn}
 \affiliation{\affA}
     \affiliation{\affB}
	\date{\today}

\begin{abstract}
Emission and absorption of energy are fundamental aspects of non-equilibrium dynamics. The heating induced by driving a many-body system is perhaps the most straightforward diagnostic of the process of equilibration, or the lack thereof. Gapless systems are particularly susceptible to drive-induced heating, and the capacity to control such heating is of experimental importance. Our study addresses this challenge in the framework of conformal field theory (CFT), for which we study families of structured drives up to the aperiodic  Thue-Morse sequence. Concretely, we consider a class of spatially inhomogeneous Hamiltonians, where the operator evolution is governed by a non-linear classical dynamical system $\mathcal{K}$. The existence of invariant regions and fixed points of $\mathcal{K}$ leads to different levels of ergodicity breaking. Upon bridging the gap between this dynamical system and the driven CFT, we classify various dynamical phases of matter, including the heating and non-heating phases, as well as a prethermal phase with a controllably slow heating rate. 
We further generalize the discussion to $\eta$-random multipolar driving, characterized by $\eta$-th order multipolar correlation in time. A ``triply tunable'' parametric dependence of the prethermal lifetime arises as $K^{-2(\eta-\xi)}$, where $K$ quantifies the deviation from the preimages of the fixed points of $\mathcal{K}$, the multipolar order $\eta$, and the order of the preimages $\xi$. 
Upon sacrificing Hermiticity by considering SU(2) deformed CFTs, we find another non-heating phase with a non-zero measure, inaccessible via purely unitary CFTs. This is underpinned by an emergent compact subspace in the generic $\mathrm{SL}(2,\mathbb{C})$\ group structure, which we also identify in the transfer matrix in non-Hermitian systems with binary disorder.

\end{abstract}
\maketitle
\let\oldaddcontentsline\addcontentsline
\renewcommand{\addcontentsline}[3]{}

\section{Introduction} 

Periodically driven systems can host novel far-from-equilibrium phenomena, absent in thermal equilibrium, such as discrete-time crystals~\cite{PhaseStructure2016,Else2016, Yao2017}. They are also versatile control tools via  Floquet engineering~\cite{schweizer2019floquet,RobustDynamic2020,Geier2021,MagnonBoundStates2023,Fu2024}. Recently, non-periodic driving without the strict temporal periodicity restriction has attracted notable research interest~\cite{Nandy2017,Dumitrescu2018,Boyers2020,Crowley2020,ZhaoRandom2021,TimmsQuantizedFloquet2021,long2021nonadiabatic,Zhao2022Suppression,Timeperiodicity2022,Moon2024,schmid2024self,Hilbert-Space-Ergodicity2024,Prethermalization2024,pilatowsky2025critically,wu2025geometricquantumdriveshyperbolically}. They allow for a variety of non-equilibrium phenomena unique to non-periodically driven systems. 
Paradigmatic examples include 
discrete-time quasi-crystals~\cite{Dumitrescu2018,zhao2019floquet,else2020long,Time-Quasicrystals2025} and time rondeau crystals~\cite{Moon2024,ma2025stabletimerondeaucrystals}, notably enriching the possible forms of temporal order in non-equilibrium settings.
  
Due to the lack of energy conservation, time-dependent systems tend to heat up to a featureless ``infinite temperature" state, where all non-trivial correlations vanish. A possible workaround for this heating issue is using strong spatial disorder, inducing the Floquet many-body localized phase~\cite{Bordia2017,Abanin2019,Decker2020} which persists at least for an exceptionally long timescale~\cite{abanin2021distinguishing}. In clean systems, Floquet heating can be parametrically suppressed in the high-frequency regime, where the driving frequency is much larger than any local energy scale in the undriven system. Hence, energy absorption from the drive normally requires multi-body resonance, which rarely occurs in locally interacting systems~\cite{ExponentiallySlowHeating2015,RigorousBound2016}. 
In fact, this intuitive argument also applies to certain non-periodically driven systems~\cite{else2020long,ZhaoRandom2021,Mori2021}. Consequently, a long-lived metastable prethermal regime appears, which can be well captured by a quasi-conserved effective Hamiltonian~\cite{Kuwahara2016Floquet-Magnus}. 

However, many physical systems do not have a well-defined local energy scale, e.g., systems at the phase transition where the correlation length diverges~\cite{francesco2012conformal,polkovnikov2011colloquium}. 
The spectrum becomes gapless at the critical point, and therefore, intuitively, these systems are particularly vulnerable to heating.  
The large correlation length also challenges the standard numerical techniques in simulating non-equilibrium gapless systems, which are inevitably restricted by notable finite-size effects. Therefore, identifying possible mechanisms to stabilize gapless systems with time-dependent drive remains a largely unexplored territory.

At the critical point, a many-body system typically becomes scale-invariant. This property implies considerable structure for the system, and a variety of analytical tools have been developed to study the critical phenomena. Among many others, conformal field theory (CFT) stands out and provides an often analytically tractable framework. It incorporates not just scale invariance but the full conformal symmetry, represented by the Virasoro algebra~\cite{francesco2012conformal}. Within the CFT framework, many exact solutions have been obtained, providing valuable insights for the entanglement entropy scaling~\cite{calabrese2004entanglement} and a wide range of critical phenomena, e.g., in the 2D Ising model~\cite{belavin1984infinite} and the Potts model~\cite{nienhuis1984critical}.

Pioneering works have focused on the quench dynamics following certain spatially inhomogeneous deformations, e.g., in the form of sine-squared deformation, where the excitation on top of a uniform initial state can be analytically captured within the CFT framework. A further, large class of inhomogeneous CFTs focuses on a so-called \slr\  deformation, where the 
Hamiltonian is written via the Virasoro
generators which form an \slr\ algebra~\cite{wen2016evolution,wen2018quantum,Lapierre2020Finestructure}. The corresponding operator evolution can be cast in terms of analytically tractable M\"obius transformations, leading to exactly solvable dynamics of the correlation function and entanglement generation.

 This framework has been  generalized beyond quenches to time-dependent systems with stepwise periodic, quasiperiodic Fibonacci and random driving protocols~\cite{berdanier2017floquet,Fan2020Emergentspatial,Lapierre2020Finestructure,ageev2021deterministic,lapierre2024floquet,das2024exactly,jiang2025new,wen2022periodically,PhysRevB.111.094304}, where 
 a sequential modulation of the spatial deformation generally induces heating with the rapid entanglement growth. 
 Accordingly, the product of M\"obius matrices, $\Pi_j=G_1G_2...G_j,$
where $G_i$ denotes the M\"obius matrix for each driving step, now governs the operator evolution and hence the long-time heating behavior.

It is within this framework that we investigate CFT models driven by a family of non-periodic yet structured protocols, including the aperiodic Thue-Morse (TM) sequence, and its truncated and randomized variant, the random multipolar driving  (RMD) protocol~\cite{ZhaoRandom2021}. By varying the driving protocols, we realize different dynamical phases of matter, e.g., the heating and non-heating phases, as well as a prethermal phase with a tunably slow heating rate. 

We use the growth rate of the entanglement entropy in the long-time limit as the order parameter to quantify heating. It
corresponds to the Lyapunov exponent $\lambda_{L}$ of $\Pi_j$, which is further bounded by the trace of $\Pi_j$. Using the self-similar structure of the TM sequence, we obtain a recursive trace map, which was first derived in the context of excitations and transport in TM quasi-crystals~\cite{axel1986vibrational,axel1989spectrum,avishai1992transmission}. 

\begin{figure}[t]
\centering
\includegraphics[width=\linewidth]{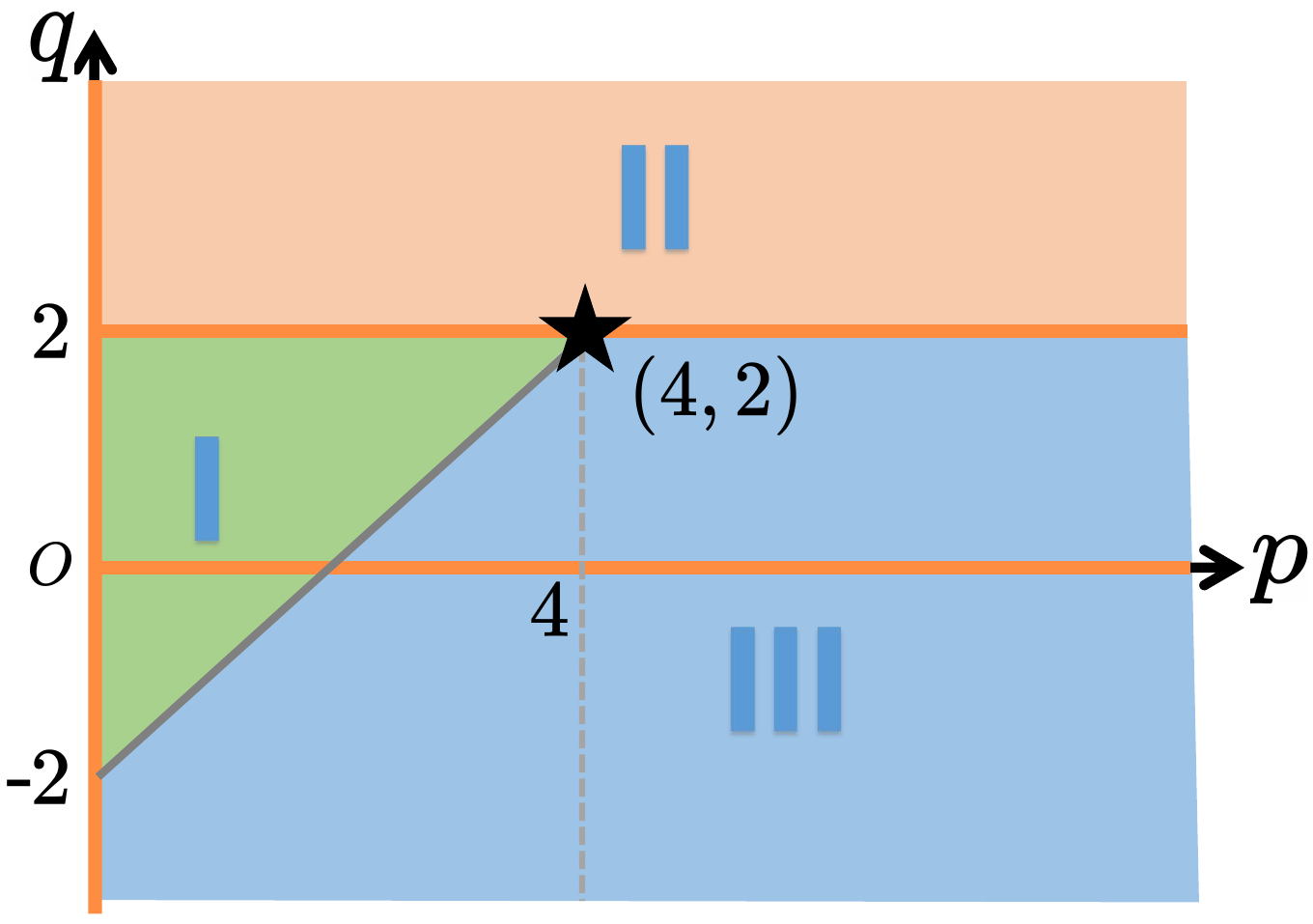}
\caption{Trace space for TM trace map showing three invariant regions, I-III. Region I is bounded, while Regions II and III are not. The gray line $p=q-2,p\in[0,4]$, as an intersection between Region I and Region III, is naturally invariant. Point $(4,2)$ labeled by a black star is a fixed point, which crucially possesses infinite preimages. For example, the three orange lines are the first three preimages of the fixed point. Higher order preimages lead to fractal fine structures, as shown in Fig.~\ref{Fig:preimageandentropy}.}
\label{Fig:TMtracemap}
\end{figure}

We recast this recursive map
as a 2D non-linear dynamical system $\mathcal{K}$, which generally exhibits chaotic dynamics with a positive Lyapunov exponent. However, crucially, the system features a set of invariant regions which map to themselves using $\mathcal{K}$, directly constraining the possible values of the Lyapunov exponent, hence breaking the ergodicity of $\mathcal{K}$. As shown in Fig.~\ref{Fig:TMtracemap}, there are three invariant regions:
\begin{itemize}
    \item Region I is compact with vanishing $\lambda_{L}$, and hence the corresponding CFT model is non-heating.
    \item Region II is non-compact, where the Lyapunov exponent is positive, resulting in heating dynamics.
    \item Region III is also non-compact, and Lyapunov exponents are generally positive. Yet, there exists an entire family of fine-structures with vanishing $\lambda_{L}$, which are obtained by identifying the preimages of the fixed point of $\mathcal{K}$.
\end{itemize}

For specific Hamiltonian parameters or driving protocols, not all regions are accessible. We first propose concrete unitary CFT models that access Region II and III. Crucially, we demonstrate that heating is strictly prohibited when the driving precisely maps onto these fine structures in Region III. Upon slightly increasing the deviation from these fine structures, the heating rate becomes finite but controllably small. It results in the long-lived prethermal phase similar to driven systems in the high-frequency regime, albeit here the stability originates from the preimages of a fixed point of the underlying dynamical system $\mathcal{K}$.

We further investigate the heating rates in systems driven by the $\eta-$RMD protocol, characterized by $\eta-$multipolar temporal correlations~\cite{ZhaoRandom2021}. It is obtained by truncating and randomizing a TM sequence. 
For $\eta=0$, the system is randomly driven by two M\"obius matrices, while the limit $\eta\to\infty$ is the TM sequence. In short-range interacting systems and for a large driving frequency $\omega$, RMD systems exhibit a characteristic ``doubly tunable'' prethermal lifetime scaling, $\omega^{2\eta+1}$~\cite{ZhaoRandom2021,Mori2021,liu2025prethermalization}. The scaling exponent may change in long-range interacting systems~\cite{Moon2024}. However, in CFT systems we observe a ``triply tunable'' parametric dependence, 
 $K^{-2(\eta-\xi)}$, which depends on the deviation $K$ away from the preimages, the multipolar order $\eta$, and the order $\xi$ of the preimages. It is worth highlighting that the extra tunability in $\xi$ stems from the analytically tractable preimages of $\mathcal{K}$, which are generally absent in generic non-integrable systems.

Finally, we go a step further and consider CFT with non-unitary evolution, where $G_i$ is now extended to \slc\ M\"obius matrices.  Non-unitary CFT have been discussed in various contexts, such as the analysis of Lee-Yang zeroes~\cite{yang1952statistical,li2023yang}, exotic entanglement scaling~\cite{chang2020entanglement} and enhanced supersymmetry~\cite{fei2015critical}. For our purpose, involving non-unitary evolution allows us to enter Region I, which is compact. This paves the way to forbid heating in a parameter space of non-zero measure. Interestingly, by designing a TM driving protocol that involves both \slr\ and SU(2) deformed CFTs, we realize a phase transition between heating and non-heating phases, with an analytically solvable phase boundary. Remarkably, exactly the same phase diagram appears for RMD protocols with $\eta\ge1$, which we account for by identifying an emergent compact subspace in the generic \slc\ group structure.

Note, in practice, physical systems generally cannot satisfy the required conformal symmetry exactly. E.g., when the system slightly deviates from the critical point, the correlation length actually becomes finite. Yet, this correlation length will still be parametrically large in the closeness to the critical point, such that our findings still provide important insights into controlling heating that occurs in regions of size much smaller than this length scale.

The paper is organized as follows. In Sec.~\ref{sec:II}, we briefly review the basic CFT framework and introduce the $\mathrm{SL}(2,\mathbb{R})$-deformed Hamiltonian.
In Sec.~\ref{sec:III}, we introduce the aperiodic Thue-Morse protocol and derive the recursive trace map. This leads to different values of the Lyapunov exponent, by which we further classify possible heating phases.
In Sec.~\ref{sec:IV}, we analyze the heating properties within Region II and Region III, and demonstrate non-heating and prethermal dynamics. 
We generalize the discussion to RMD and non-unitary CFT systems in Sec.~\ref{Sec:RMD} and Sec.~\ref{sec:VI}, respectively.
Finally, we give a conclusion and outlook in sec.~\ref{sec:VII}.

\label{sec:classification}

\section{\slr\ deformed Hamiltonian in Driven CFT}\label{sec:II}


We start by introducing the \slr\ deformed Hamiltonian, which can be realized in lattice models at the critical point, eg, XXZ chain and free fermion systems~\cite{wen2018quantum,lapierre2020emergent}(also see Appendix.~\ref{Appendix:free}).  We then show that, in such driven systems, the Heisenberg evolution of operators reduces to the product of a series of matrices with SU(1,1) symmetry, which is isomorphic to \slr~\cite{2018arXiv180500031W,2022arXiv221100040W,Fan2020Emergentspatial,PhysRevResearch.3.023044,choo2022thermal,lapierre2025driven}.


For simplicity, we focus on the time evolution generated by piecewise constant Hamiltonians. Starting from an initial state $|\psi_0\rangle$, the system evolves as 
\begin{eqnarray}
\label{eq:time-evolution}
        |\psi_j\rangle=\hat U(t)|\psi_0\rangle,  \hat U(t)=\prod_{l=1}^j e^{-iH^{(l)} T^{(l)}},
\end{eqnarray}
 where each Hamiltonian $H^{(l)}$ acts on the system for time duration $T^{(l)}$ and the total time reads $t{=}\sum_{l=1} ^jT^{(l)}$.  Correspondingly, the operator evolution becomes $\langle \hat{\mathcal{O}}(x,t)\rangle{=}\langle \psi_0|\hat U^{\dagger}(t) \hat{\mathcal{O}}(x,0)\hat U(t)|\psi_0\rangle$.

The initial state $|\psi_0\rangle$ is chosen to be the ground state of a uniform Hamiltonian, $   H_0=\frac{1}{2\pi}\int_0^L[T(x)+\bar T(x)]dx$, where $T(x) $ and $\bar T(x)$ are the chiral and antichiral (holomorphic and anti-holomorphic) energy momentum
tensor with a translation symmetry, respectively(see Appendix.~\ref{Appendix:intro to CFT} and Refs.~\cite{ginsparg1988applied,francesco2012conformal} for further details). $L$ denotes the size of the system. 

The \slr\ spatial deformation Hamiltonians are defined as 
$H{=}\frac{1}{2\pi}\int_0^L[f_r(x)T(x)+g_r(x)\bar T(x)]dx,$
where $f_r(x)$ and $g_r(x)$ are real functions that define the spatial deformation profile~\cite{PhysRevResearch.3.023044}. We assume periodic boundary conditions unless stated otherwise~\footnote{When $r=1$, we consider open boundary conditions since the ground state of $H_0$ will be annihilated by the deformed  Hamiltonian $H$ with the periodic boundary~\cite{2018arXiv180500031W,wen2022periodically}. }. For the deformation to be an \slr, $f_r(x)$ needs to take the following form
\begin{align}
f_r(x)=\sigma^0+\sigma^+\cos\frac{2\pi  r x}{L}+\sigma^-\sin\frac{2\pi  rx}{L},r\in \mathbb{Z},\label{Eq:Hamiltonian}
\end{align}
where $\sigma^0,\sigma^+,\sigma^-{\in} \mathbb{R}$. The same condition holds for $g_r(x)$~\footnote{With this deformation, the Hamiltonian $H$ can be written as a real linear combination of the generator of SL$^{(r)}(2,\mathbb{R})$, which is isomorphic to an $r$ fold cover of \slr\ group~\cite{wen2022periodically}. }.  We note that the deformation can be generalized to an \slc\  one with $f_r(x)$ and $g_r(x)$ being complex, as discussed in Sec.~\ref{sec:nonhermitian-heating}. 

This special deformation enables analytical simplification for the operator evolution~\cite{wen2018quantum}. This is achieved by first transforming the Lorentzian spacetime $(x,t)$ into Euclidean spacetime $(x,\tau=it)$, and then defining the complex coordinates (holomorphic and anti-holomorphic) $\omega=\tau+ix,\bar\omega=\tau-ix$.
Then, we map $(\omega,\bar \omega)$ to an $r$-fold
complex plane by defining $z=\exp(2\pi\omega /l),\omega=\tau+ix,l=L/r$.  On the complex plane, the time evolution of a primary field $\hat{O}(z_1,\bar z_1)=e^{iHt}\hat{O}(z,\bar z)e^{-iHt}$ is determined by a conformal mapping $(z,\bar z)\rightarrow(z_1,\bar z_1)$ through 
 $\hat{O}(z_1,\bar z_1)=(\frac{\partial z_1}{\partial z})^h (\frac{\partial \bar z_1}{\partial z})^{\bar h} \mathcal{O}(z,\bar z)$, where $h,\bar h$ denote the conformal dimension of $\mathcal{O}(z,\bar z)$. 
 
 Notably, for the deformed Hamiltonian Eq.~\eqref{Eq:Hamiltonian}, the conformal mapping is particularly simple and can be cast as a M\"obius transformation, $z_1=(az+b)/(cz+d)$. We can also define the M\"obius matrix  $$G\equiv \begin{pmatrix}
 a & b\\
 c & d
\end{pmatrix},$$ which belongs to SU(1,1) group.

The concrete form of the above holomorphic M\"obius matrix depends on both the time duration and specific parameters, e.g., $f_r(x)$, of the driving Hamiltonian.  
For example, the uniform Hamiltonian $H_0$ generates an M\"obius matrix 
\begin{align}
    U_0(T)=\begin{pmatrix}
  e^{ i \pi \frac{T}{L}}& \\
  &e^{- i \pi \frac{T}{L}}
\end{pmatrix},\label{Eq:U0T}
\end{align}for a time duration $T$.
For Hamiltonian $H_1$ with spatial deformation $\sigma^0=\sigma^+=1$, the M\"obius matrix becomes 
\begin{align}
    U_1(T)=\begin{pmatrix}
 1+i\frac{\pi T}{l}  &-i\frac{\pi T}{l}  \\
 i\frac{\pi T}{l}  &1-i\frac{\pi T}{l} 
\end{pmatrix}. \label{Eq:U1T}
\end{align}

Therefore, the operator time evolution in Eq.~\eqref{eq:time-evolution} is now determined by the product of M\"obius matrices 
\begin{align}
 \Pi_j=G_1G_2...G_j\equiv\begin{pmatrix}
 \alpha_j&\beta_j\\
\gamma_j  &\delta_j
\end{pmatrix}.\label{Eq:productG}
\end{align}
The anti-holomorphic part can also be derived in a similar manner $$\tilde \Pi_j{\equiv}\begin{pmatrix}
 \alpha_j'&\beta_j'\\
\gamma_j' &\delta_j'
\end{pmatrix}.$$
For simplicity, in the case of \slr\ 
 deformation, we consider $f_r(x)=g_r(x)$, so that the holomorphic and anti-holomorphic M\"obius matrices are the same.

We use the time-dependence of the bipartite entanglement entropy as a diagnostic of the heating process. Within the CFT framework, this can be efficiently calculated. We consider the reduced density matrix $\hat{\rho}_A(j)$ for the subregion $A = [x_1 = -\frac{1}{2}l, x_2 = \frac{1}{2}l]$, which is given by $\hat{\rho}_A(j) = \mathrm{tr}_{\bar{A}}(|\psi_j\rangle\langle\psi_j|)$, where $\bar{A}$ denotes the complement of $A$. The $m$-th order Renyi entropy reads \begin{eqnarray}
    \begin{aligned}
        S_m(j)&\equiv \frac{1}{1-m}\ln [\mathrm{tr}(\hat \rho_A^m(j))]\\&=\frac{1}{1-m}\ln \langle \psi_j|\mathcal{T}_m(x_1,0)\mathcal{T}_m(x_2,0)|\psi_j\rangle,
    \end{aligned}
\end{eqnarray} where $\mathcal{T}_m$ denotes the twist field operator with conformal dimension $h_m=\bar h_m=\frac{c}{24}(m-1/m)$ and central charge $c$. 
By taking the limit $m\rightarrow 1$, we obtain the following general form for the entanglement entropy
\begin{eqnarray}
    \begin{aligned}
    S_A(j)-S_A(0)&=\frac{c}{6}\ln \left( \alpha_j \gamma_j -\alpha_j\delta_j-\gamma_j\beta_j
+\beta_j\delta_j\right)\\
+ &\frac{c}{6}\ln \left( \alpha'_j \gamma'_j -\alpha'_j\delta'_j-\gamma'_j\beta'_j
+\beta'_j\delta'_j\right).\label{Eq:SA}
\end{aligned}
\end{eqnarray}
In fact, when $\Pi_j$ and $\tilde \Pi_j$ are both SU(1,1) M\"obius matrices, Eq.~\eqref{Eq:SA} simplifies to 
\begin{eqnarray}
    S_A(j)-S_A(0)=\frac{c}{3}(\ln |\alpha_j-\beta_j|+\ln |\alpha'_j-\beta'_j|).
    \label{eq.entanglementSU11}
\end{eqnarray}
where $\delta_j=\alpha_j^*,\gamma_j=\beta_j^*,\delta_j'=(\alpha_j')^*,\gamma_j'=(\beta_j')^*$ for SU(1,1) matrices.

In the heating phase, one expects the entanglement entropy to increase linearly in the number of driving steps $j$ at a rate $\gamma$. By contrast, in the non-heating phase, $S_A(j)$ stays close to its initial value. In the long time limit, $\gamma$ reduces to $2c\lambda_L/3$(see Appendix~\ref{part:proofeescaling})~\cite{wen2022periodically}, where $\lambda_L$ corresponds to the Lyapunov exponent defined by
$$ \lambda_L\equiv \lim\limits_{j\rightarrow \infty} 
{\ln ||\Pi_j||_F}/{j}, $$ with $||...||_F$ being the Frobenius  norm of a matrix.

For M\"obius matrices considered in this work, $\lambda_{L}$ is bounded by the trace of $\Pi_j$, since $||\Pi_j||_F{\ge}  |\mathrm{tr}(\Pi_j)|/\sqrt{2}$. This trace turns out to be more analytically tractable, which we use in the following to systematically classify different heating phases of matter.

\begin{table*}[ht]
  \centering
  \renewcommand{\arraystretch}{1.5}
  \begin{tabular}{|c|c|c|c|}
    \hline
    $\sigma^0$ & $\sigma^+$ & $\sigma^-$ & M\"obius matrices  \\
    \hline
    1 & 0 & 0 &
    $\displaystyle
      U_0(T)=\begin{pmatrix}
        e^{ i \pi T/L} & 0\\
        0 & e^{- i \pi T/L}
      \end{pmatrix}$ \\ \hline
    1 & 1 & 0 &
    $\displaystyle
      U_1(T)=\begin{pmatrix}
        1+i\pi T/l & -i\pi T/l\\
        i\pi T/l   & 1-i\pi T/l
      \end{pmatrix}$ \\ \hline
    0 & 0 & 1 &
    $\displaystyle
      U_2(T)=\begin{pmatrix}
        \cosh(\pi T/l) & i\sinh(\pi T/l)\\
        -i\sinh(\pi T/l) & \cosh(\pi T/l)
      \end{pmatrix}$ \\ \hline
    $\cos\Gamma$ & $i\sin\Gamma$ & 0 &
    $\displaystyle
      U_3(T)=\begin{pmatrix}
        \cos(\pi T/l)+i\cos\Gamma\sin(\pi T/l) & -\sin(\pi T/l)\sin\Gamma\\
        \sin(\pi T/l)\sin\Gamma & \cos(\pi T/l)-i\cos\Gamma\sin(\pi T/l)
      \end{pmatrix}$ \\ \hline
  \end{tabular}
  \caption{Operator time evolution is governed by 
  M\"obius matrices in CFT systems. We illustrate  M\"obius matrices for different Hamiltonian parameters $\sigma^0$, $\sigma^+$, and $\sigma^-$. The first three M\"obius matrices, $U_{0/1/2}$ belong to \slr\ (see Sec.~\ref{sec:II}), whereas $U_{3}$ belongs to SU(2) (see Sec.~\ref{sec:VI}). }
  \label{tab:mobius}
\end{table*}

\section{{Thue-Morse protocol, trace maps and invariant subsets}}\label{sec:III}

\begin{figure}[t]
    \centering
    \includegraphics[width=\linewidth]{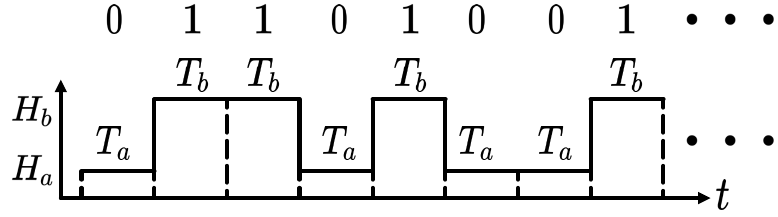}
    \caption{Schematic of the Thue-Morse driving protocol, where the temporal modulation follows the binary sequence shown in the top row. ``0" corresponds to applying the Hamiltonian $H_a$ for a duration of $T_a$, while ``1" corresponds to applying $H_b$ for a duration of $T_b$. }
    \label{fig:TMdrive}
\end{figure}

We start by introducing the aperiodic TM sequence involving two elementary letters, ``0" and ``1", which can be constructed through a substitution rule. Starting with 0, we repetitively replace $0{\rightarrow} 01, 1{\rightarrow} 10$, and finally obtain an infinite aperiodic sequence $0110 \, 1001\,...$, as shown in Fig.~\ref{fig:TMdrive}.  The set $\{H_{a(b)},T_{a(b)}\}$ defines a set of M\"obius matrices $\{\mathcal{M}_0, \mathcal{N}_0\}$, e.g., chosen from Table \ref{tab:mobius},  also forming a TM sequence~\cite{Mori2021} . The corresponding operator evolution can be well described by the recursive relation
\begin{eqnarray}
\begin{aligned}
   & \mathcal{M}_{1}=\mathcal{M} _0\mathcal{N}_0,\mathcal{N}_1 =\mathcal{N}_0\mathcal{M} _0, ... \\ 
       & \mathcal{M}_{n}=\mathcal{M} _{n-1}\mathcal{N}_{n-1},\mathcal{N}_n =\mathcal{N}_{n-1}\mathcal{M} _{n-1}.\label{Eq:recursiverelation}
\end{aligned}
\end{eqnarray}
Hence, the number of M\"obius matrices $j$ grows exponentially in the stroboscopic time $n$, $j=2^n$. 

Through the iteration rule in Eq.~\eqref{Eq:recursiverelation}, one obtains the trace map~\cite{axel1986vibrational,axel1989spectrum,avishai1992transmission}
\begin{eqnarray}
    \begin{aligned}
        x_{n+1}=x_{n-1}^2(x_n-2)+2,n\ge 2,\label{Eq:TM}
\end{aligned}
\end{eqnarray}
where  $x_n=\mathrm{tr}(\mathcal{M} _{n})$ and we use the uni-modular nature of $\mathcal{M}_{n}$. 
This trace map starts from initial conditions $x_0$ and $x_1$, which can be determined by the driving parameters, e.g., $T_0$ and $T_1$. It allows us to efficiently obtain the asymptotic solution of $\mathrm{tr}(\mathcal{M}_{n})$ in the limit $n{\to}\infty$ and determine the Lyapunov exponent $\lambda_L$. Note, Eq.~\eqref{Eq:TM} generally holds for complex numbers, but we will restrict it to real numbers when $\mathcal{M}_{n}$ is \slr. 

Generic discrete non-linear maps generate chaotic trajectories in phase space, leading to positive Lyapunov exponents. However, the system features three invariant regions that map to themselves,  
which constrain the asymptotic behavior and break ergodicity of the non-linear system.
To see this, we define  new coordinates $p_n=x_n^2,q_n=x_{n+1}$ and convert Eq.~\eqref{Eq:TM} to a periodic 2D dynamical system, 
$(p_n,q_n)=\mathcal{K}(p_{n-1},q_{n-1})$ 
with 
\begin{eqnarray}
    \begin{aligned}
    \mathcal{K}(p,q)=(q^2,pq-2p+2).
    \label{Eq:tracemap2d}
\end{aligned}
\end{eqnarray}
According to Refs.~\cite{axel1986vibrational,1989JSP....57.1013A,avishai1992transmission}, this 2D plane can be divided into three subsets as shown in Fig.~\ref{Fig:TMtracemap},
\begin{itemize}
    \item Region I=$\{(p,q)|p\ge  0, p-2\le q\le 2 \}$,
    \item  Region II=$\{(p,q)|p\ge  0, q\ge 2 \}$,
    \item Region III=$\{(p,q)|p\ge  0,  q\le 2, p-2\ge q \}$,
\end{itemize}
which are invariant under the dynamical map Eq.~\eqref{Eq:tracemap2d}:
given one initial condition $(p_1,q_1)$, trajectories generated by $\mathcal{K}$ are constrained to one of those regions.

Region I is compact and hence $(p_n,q_n)$ therein always remains finite. This implies a vanishing Lyapunov exponent. All other regions are non-compact, allowing the non-linear map $\mathcal{K}$, $(p_n,q_n)$ to generate an evolution towards infinity, typically resulting in a positive Lyapunov exponent. A non-vanishing Lyapunov exponent is indeed found for all points in Region II and almost all points in Region III. 
However, Region III features some fractal fine-structures, forbidding  the unbounded divergence at long times. 

To see this, note that within Region III,
Eq.~\eqref{Eq:tracemap2d} has one fixed point $(p_f,q_f)=(4,2)$.  
We can determine preimages of this fixed point by iteratively solving the inverse map $\mathcal{K}^{-1}$. For convenience, we dub the preimages obtained by applying $\mathcal{K}^{-1}$ iteratively $\xi$ times as the $\xi$-th order preimages. The solution can be analytically obtained for small $\xi$, as denoted by the three orange lines in Fig.~\ref{Fig:TMtracemap} for $\xi=1,2,3$. However, the structure of preimages becomes more complex as $\xi$ grows. As shown in Fig.~\ref{Fig:preimageandentropy} (a) and (b), higher-order preimages (black lines) exist in both Region I and III with zero measure. Interestingly, it was suggested in Refs.~\cite{1989JSP....57.1013A, PhysRevB.43.1034} that preimages form a {fractal} structure. 

Note, by construction, these preimages naturally correspond to initial conditions, or driving parameters, that result in zero Lyapunov exponent, since they evolve back to the fixed point after a few iterations. Now we are ready to use this framework to identify various dynamical phases in aperiodically driven CFTs.

It is worth noting that for specific Hamiltonian parameters or driving protocols, not all invariant regions are accessible. In Sec.~\ref{sec:RegionIII} we discuss one TM driven \slr\ deformed Hamiltonian where the initial condition $(p_1,q_1)$ is restricted in Region III. Later in Sec.~\ref{sec:nonhermitian-heating}, we consider CFT with non-unitary evolution such that Regions I and II become accessible.

\section{Non-heating and prethermal phases in Region III}\label{sec:IV}
\label{sec:RegionIII}
\begin{figure}[t]
\centering
\includegraphics[width=\linewidth]{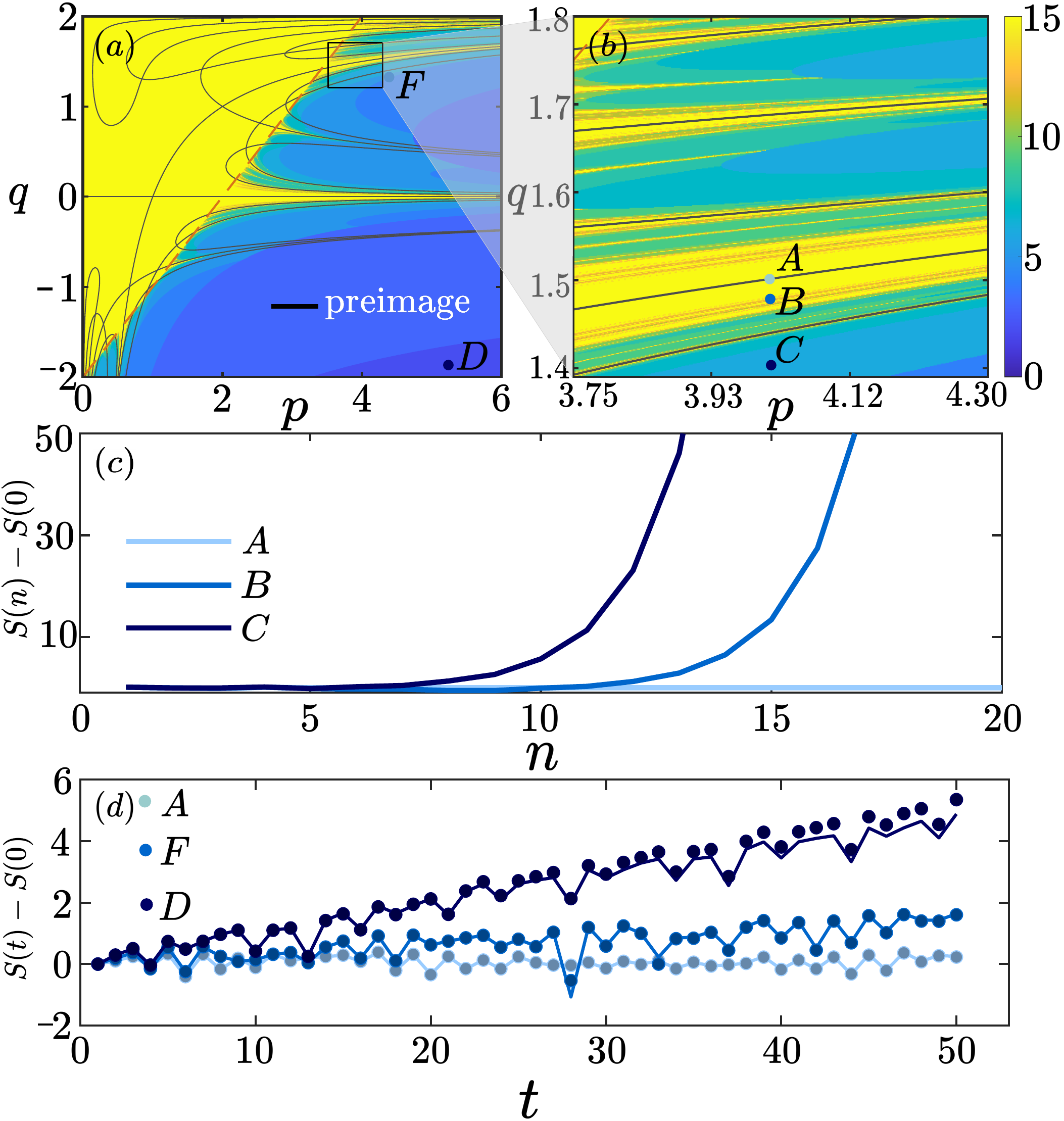}
\caption{Heating dynamics in \slr\ CFTs driven by a Thue-Morse sequence. (a) Heating time in the 2D parameter space. Black lines correspond to the preimages of a fixed point, obtained by solving the inverse map $\mathcal{K}^{-1}$ 8 times.  By construction, preimages have a zero Lyapunov exponent.
Bright areas correspond to slow heating, concentrated around the preimages. (b) Heating time in a smaller parameter space with clearer fine structures.  (c) Entanglement growth versus  stroboscopic time $n$ for different driving parameters. The heating rate is significantly suppressed for parameters close to the preimages. (d) Free fermion numerical verification(dots) of the CFT prediction(solid line). We use $L=3000$, $r=1$ and parameters $T_0,T_1$ are determined by the position of initial points. Coordinates of sampled points: A $(4,1.5)$; B $(4,1.45)$; C $(4,1.4)$; D $(5.9,-1.8)$; F $(4.8,1.2)$. 
}
\label{Fig:preimageandentropy}
\end{figure}

We consider the following driving M\"obius matrices $\mathcal{M}_0=U_0(T_0), \mathcal{N}_0=U_1(T_1)$ defined in Eq.~\eqref{Eq:U0T} and Eq.~\eqref{Eq:U1T} with $r{=}1$ in Eq.~\eqref{Eq:Hamiltonian}. As shown in Refs.~\cite{2003math......5096G,PhysRevB.43.1034}, a generic SU(1,1) matrix cannot have a trace in Region I. Indeed, by explicitly calculating the initial condition for our CFT model 
\begin{align}\label{Eq:initalconditionparameter}
    p_1&=[\mathrm{tr}(\mathcal{M}_1)]^2=4[\cos(\pi \frac{T_0}{L})-\pi \frac{T_1}{L}\sin(\pi \frac{T_0}{L})]^2,\\\nonumber
   q_1&=\mathrm{tr}(\mathcal{M} _2)=2[\cos(2\pi \frac{T_0}{L})-2\pi \frac{T_1}{L}\sin(2\pi \frac{T_0}{L})],
\end{align}
one can show that the distance between $(q_1,p_1)$ and the grey line in Fig.~\ref{Fig:TMtracemap}, 
\begin{eqnarray}
    d_n{\equiv} p_n{-}q_n{-}2=4\pi^2 (\frac{T_1}{L})^2 \sin^2(\frac{\pi T_0}{L}),
\end{eqnarray} is always positive.  

We consider the half-chain entanglement entropy in systems with open boundary conditions, and Eq.~\eqref{Eq:SA} reduces to $S_A(j)-S_A(0)=\frac{c}{6}\ln |\alpha_j-\beta_j|+ \text{antichiral\ part}$. It differs from Eq.~\eqref{eq.entanglementSU11} by a prefactor $2$ due to the boundary condition. For suitable parameters $T_0/L$ and $T_1/L$, we have $(p_1,q_1){=}(4,1.5)$, point A in Fig.~\ref{Fig:preimageandentropy}(b). It is located exactly on the preimage, black lines in Fig.~\ref{Fig:preimageandentropy}(b) in Region III. Consequently, the system remains non-heating, as verified in Fig.~\ref{Fig:preimageandentropy}(c), where the entanglement entropy (light blue) stays close to the initial value throughout the entire evolution. In contrast, driving parameters away from preimages, e.g., points C in Fig.~\ref{Fig:preimageandentropy} (b), generally lead to heating dynamics, dark blue line in Fig.~\ref{Fig:preimageandentropy} (c). 

Crucially, the time scales required for the onset of heating can be systematically prolonged by reducing the deviation from the preimages. For instance, as shown in Fig.~\ref{Fig:preimageandentropy}(c), growth of the entanglement entropy for point B is notably smaller than point C. Note, since we use stroboscopic iteration time $n$ in Fig.~\ref{Fig:preimageandentropy}(c), heating times for points B and C indeed differ by one order of magnitude in terms of the number of M\"obius matrices ($2^{n}$) being applied.

We extract the heating time $n^*$ by counting the number of stroboscopic iterations required for the trajectory to escape from a predefined finite area ($|q|<50,|p|<2500$) in the TM trace space. Fig.~\ref{Fig:preimageandentropy} (a) visualizes the heating times, and most regions are indeed blue (small $n^*$), indicating that within a few steps the system quickly heats up. Clearly, bright yellow areas concentrate around the preimages, which become particularly manifest in Fig.~\ref{Fig:preimageandentropy} (b). 

Phenomenologically, this slow-heating behavior is reminiscent of the prethermal phenomenon observed in high-frequency driven systems, although the underlying mechanism is fundamentally different. In fact, our CFT systems are far from any high-frequency limit as we choose the parameters $T_0/L$ and $T_1/L$ to be finite and $T_{0/1}$ scale up with the system size. 
Note, one can also fix $T_{0/1}$ to be a constant, but the system will reduce to a trivial quench setup in the limit $L\to\infty$.
Here, the existence of the preimages of the fixed point guarantees no heating absorption from the drive, even though the underlying system is gapless. It further underpins the prethermal phenomenon in the vicinity of the preimages, which now has a finite measure.

This becomes particularly important in detecting this phenomenon in physical systems, considering the fact that many perturbative processes can violate the ideal conformal symmetry. 
A case in point is the critical free-fermion lattice systems, whose low-energy properties are well captured by our CFT model, but driving-induced high-energy excitations inevitably violate the conformal symmetry. As shown in Fig.~\ref{Fig:preimageandentropy}, for Point A in panel (b), free fermion simulations (light blue dots) closely follow the CFT prediction (light blue line) and heating is entirely forbidden. Point F deviates from the exact preimage slightly, resulting in slowly increasing entanglement. As for Point D, heating happens considerably more rapidly and the free fermion simulation exhibits a noticeable deviation from the CFT prediction around $t\sim40$. Therefore, despite the fact that conformal symmetry is not perfect, the underlying heating suppression mechanism still plays a crucial role in stabilizing the system.

We have also analyzed Fibonacci-driven CFTs, i.e., the sequence of $\mathcal{M}_0$ and $\mathcal{N}_0$ following the Fibonacci sequence~\cite{PhysRevResearch.3.023044}. The Möbius matrices also obey a trace map $\mathcal{F}$, which now exhibits a strict invariant that constrains the asymptotic behavior\cite{PhysRevResearch.3.023044,Lapierre2020Finestructure}. At the fixed point of $\mathcal{F}$, the system is equivalent to a quenched system where driving is effectively absent. As a consequence of this invariant, the fixed point does not have any non-trivial preimages that can forbid the divergence of the trace at long times. Therefore, one cannot identify non-heating regimes that are analytically solvable as in the TM system.

\section{Random multipolar driving}\label{Sec:RMD}

The existence of fixed points and their preimages provides exceptional stability of CFT systems driven by the TM protocol.
One natural question is how robust the stability is when the driving protocol involves temporal randomness. Naively, one would expect that the system heats up swiftly as the recursive trace map no longer exists. Indeed, this is true if the driving protocol is purely random, as seen in Ref.~\cite{wen2022periodically}. However, we will show that, by considering the $\eta$-random multipolar driving ($\eta$-RMD) with a tunable degree of temporal correlations~\cite{ZhaoRandom2021}, the heating rate becomes ``triply tunable'', leading to long-lived prethermal phenomenon
even when temporal randomness is present.

$\eta$-RMD is constructed by truncating the recursive construction Eq.~\eqref{Eq:recursiverelation} at a finite order $\eta$ and generating a random sequence composed of two elementary building blocks, $\mathcal{M}_{\eta}$ and $\mathcal{N}_{\eta}$. When $\eta=0$, this corresponds to a random drive, while the limit $\eta\rightarrow\infty$ corresponds to the TM driving.
This driving protocol naturally separates driving parameters into two classes, depending on whether they can be mapped back to a region in Fig~\ref{Fig:TMtracemap} that is sufficiently close to the fixed point within the $\eta$-step iteration of the map $\mathcal{K}$. These two situations exhibit notably different heating behaviors, which we account for via a perturbative analysis.
For simplicity, we first consider driving parameters around the fixed point, then generalize the analysis to the other higher-order preimages.

\subsection{Fixed point}
\label{sec.fixedpoint}
\begin{figure}[t]
\centering
\includegraphics[width=1\linewidth]{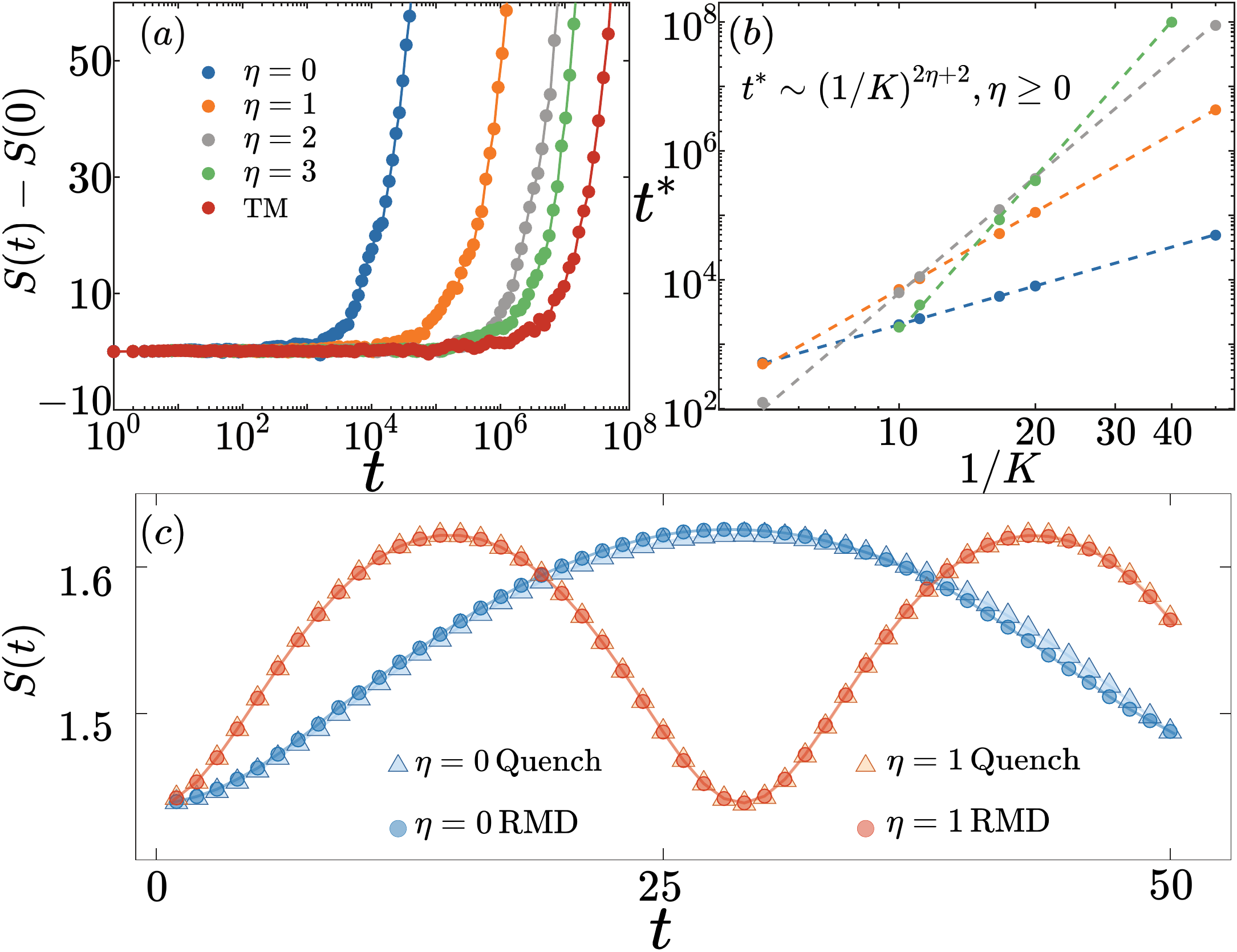}
\caption{(a) Entanglement entropy growth for $\eta$-RMD and TM drive with $K=0.05,\ell_1=0$ for driving parameters close to the fixed point. The prethermal lifetime increases for larger $\eta$, highlighting the importance of temporal correlation in stabilizing the system. (b) Dependence of the prethermal lifetime $t^*$ on $K^{-1}$ in a log-log scale. The slope of the power law dependence strongly depends on the multipolar order as $2\eta+2$.
(c) Free fermion numerical comparison of the RMD driving (circles) and quench dynamics (triangles) generated by the effective Hamiltonian \( H_{\text{eff}} \). The effective Hamiltonian accurately captures the early time evolution of the RMD system. We use \(\sigma_0 = 1\), \(\sigma^+ = 1/2\) and \(\sigma_0 = 2\), \(\sigma^+ = 1\) for $\eta=0$ and $\eta=1$, respectively. We also 
use \({T}{L^{-1}} = K\), $K=0.02$, $\ell_1=0$ and $L=1000$ for the simulation.}
\label{Fig:rmddiffnentropy}
\end{figure}
Before we give a more detailed analysis, it is worth clarifying that the $(p,q)$ space (Fig.~\ref{Fig:TMtracemap}) and trajectories therein are not well-defined here in a strict sense, due to the truncation of the trace map Eq.~\eqref{Eq:TM}  at a finite value, $n=\eta$. However, they can still be useful in designing the driving protocol. We parametrize the distance $K$ between the initial condition $\left (\mathrm{tr}(\mathcal{M}_1)^2,\mathrm{tr}(\mathcal{M}_2) \right )$  and the fixed point, or preimages in the next section, as a tuning parameter of the heating rate.

Concretely, we consider the driving parameters $T_0/L=\ell_1+K,T_1/L=K,\ell_1\in \mathbb{Z}, K\in \mathbb{R}$, such that for small $K$ the distance between {$\left (\mathrm{tr}(\mathcal{M}_1)^2,\mathrm{tr}(\mathcal{M}_2) \right )$} and the fixed point, $(p=4,q=2)$ in Fig.~\ref{Fig:TMtracemap}, scales as $K^2$. 

We calculate the entanglement entropy using Eq.~\eqref{Eq:SA}. The results are shown in Fig.~\ref{Fig:rmddiffnentropy}(a) for different $\eta$ and a fixed $K=0.05$. The entanglement entropy remains close to zero at short times before the notable increases at longer times. 

The prethermal lifetime $t^*$ strongly depends on the
multipolar order $\eta$.
As $\eta$ increases, $t^*$ grows and approaches the TM limit (red), highlighting the importance of the temporal correlations in stabilizing a randomly driven CFT system.  

To further quantify their dependence, we extract the prethermal lifetime as the time $t^*$ when the entanglement growth first exceeds a threshold $S^*$. In our numerical simulation, we choose $S^*=10$ but our findings are independent of the specific choice of $S^*$. Fig.~\ref{Fig:rmddiffnentropy}(b) shows the dependence of $t^*$ on $1/K$, which exhibits the algebraic scaling 
\begin{equation}
        t^*\sim \left ({K}\right )^{-2\eta-2}, \eta\ge0
\label{Eq:fixedscaling},
\end{equation}
for large $K^{-1}$ \footnote{$\eta=0$ reduces to the purely randomly drive, and the scaling $t^*\sim (1/K)^2$ also matches with the result reported in Ref.~\cite{wen2022periodically}}.
In generic many-body systems with local interactions in the high-frequency regime, we have found $t^*\sim \omega^{2\eta+1}$~\cite{ZhaoRandom2021}. In contrast, we do not have a natural local energy scale in CFT systems, yet we still observe the characteristic scaling with an $\eta-$dependent exponent.
 
\begin{figure}[t]
\centering
\includegraphics[width=1\linewidth]{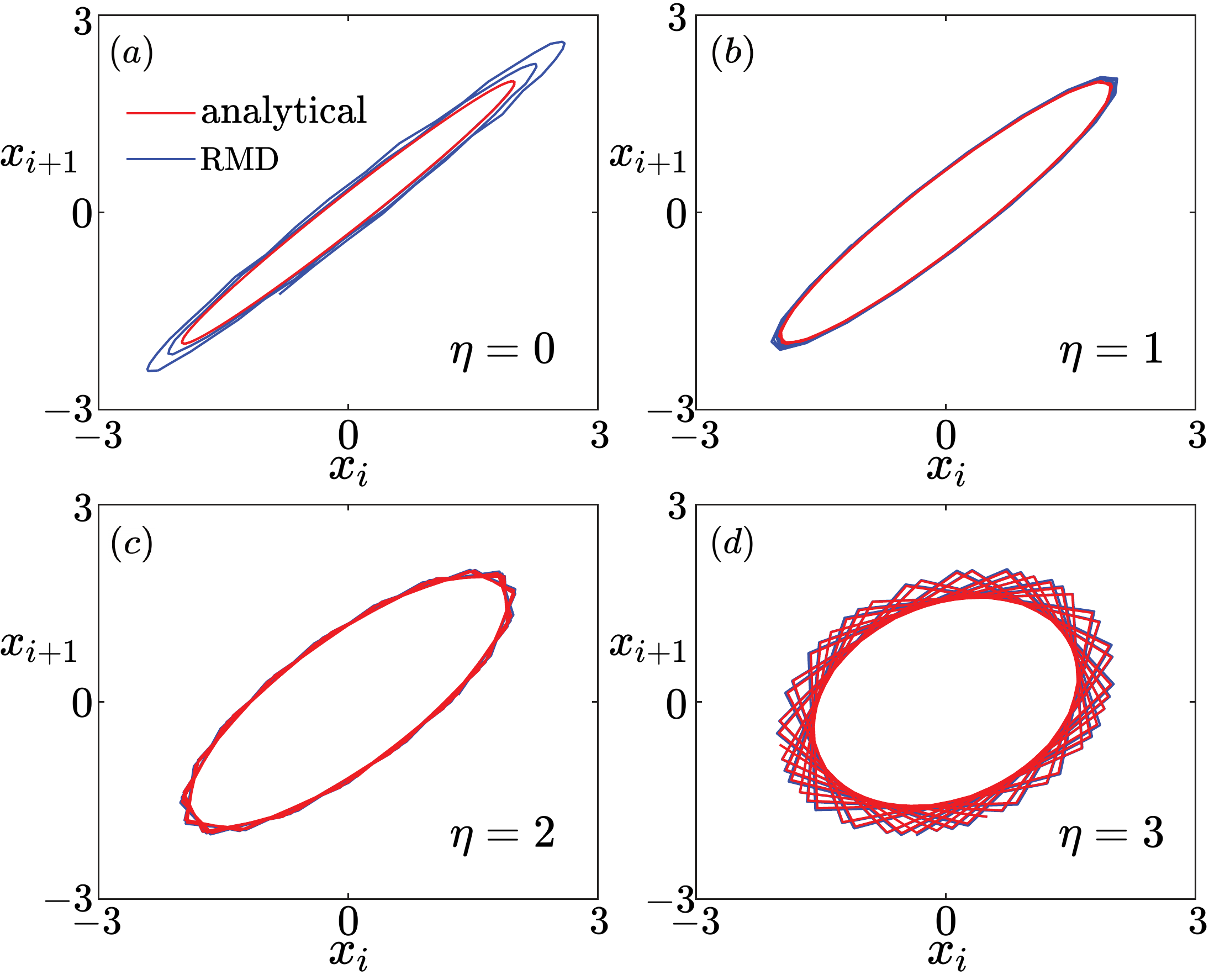}
\caption{Trace trajectory of $\eta$-RMD for driving parameters around the fixed point. The red curves denote the analytical trajectories derived from the quenched dynamics generated by $\bar{M}_{\eta}'$, while the blue curves denote the RMD trajectories. As $\eta$  increases, the agreement between the two improves.
Here we set $\ell_1=0,K=0.04$.}
\label{Fig:fixedpointtra}
\end{figure}

We also analyze the evolution of the trace of the M\"obius matrix, \( x_i = \mathrm{tr}(\Pi_i) \), where $\Pi_i$ is the product of a random selection of $\mathcal{M}_{\eta}$ and $\mathcal{N}_{\eta}$. Due to the lack of self-similarity of the driving protocol, the recursive trace map, e.g., Eq.~\eqref{Eq:TM}, is not applicable. 
The blue curves in Fig.~\ref{Fig:fixedpointtra} depict trace trajectories for different $\eta$-RMDs. For short times, the trajectory oscillates and forms a circular orbit. However, at longer times, the trajectories begin to spiral outward stochastically, and the inverse of the growth rate of the radius also follows Eq.~\eqref{Eq:fixedscaling}.

This behavior can be perturbatively justified by investigating the stochastic process 
induced by
\begin{align}
     \bar M_{\eta}=\frac{1}{2}(\mathcal{M}_{\eta}+\mathcal{N}_{\eta}), D_{\eta}=\frac{1}{2}(\mathcal{M}_{\eta}-\mathcal{N}_{\eta}),
 \end{align}
such that
$\mathcal{M}_{\eta}$ and $\mathcal{N}_{\eta}$ correspond to $\bar M_{\eta}+\varsigma D_{\eta}$, where $\varsigma $ is a random variable, being either $+1$ or $-1$ with equal probability. The average matrix $\bar M_{\eta}$ is not area-preserving in the 
trace space since its determinant is larger than 1. For small $K$, one can perturbatively obtain the leading order contribution
\begin{align}
    \det (\bar M_{\eta})\approx1+\gamma K^{2{\eta}+2}, {\eta}\ge 0,
\end{align}
with a positive coefficient $\gamma$ \footnote{Here we only consider the case with $\ell_1\in \text{even}$ while the odd case can be reduced to the even case with a global $\pi$ phase.
}. One can normalize the average matrix as $\bar M_{\eta}'=\bar M_{\eta}/\sqrt{\det(\bar M_{\eta})}$, such that $\bar M_{\eta}'$ becomes area-preserving. It leads to the trace trajectory 
\begin{align}
\label{eq.closedorbit}
(x_i,x_{i+1})=\left(\cos(2i\theta\right), \cos(2(i+1)\theta),
\end{align}
where $\theta = \arccos(\mathrm{tr}(\bar M_{\eta}') / 2)$ determines the oscillation frequency. We plot these trajectories (red) in Fig.~\ref{Fig:fixedpointtra} for different $\eta$, which accurately capture the early time oscillations as seen in the RMD trajectories (blue). In fact, as shown in Fig.~\ref{Fig:rmddiffnentropy}(c), clear coherent oscillations in entanglement dynamics also appear in the numerical simulations of free fermion dynamics generated by RMD protocols(circles). 

We also construct an effective Hamiltonian $H_{\mathrm{eff}}$ by comparing the normalized matrix, $\bar{M}_{\eta}'$ to a general $\mathrm{SU}(1,1)$ matrix (parametrized by $\sigma^0$ and $\sigma^{\pm}$ in Eq.~\eqref{Eq:Hamiltonian}), to capture this early time dynamics. As shown in Fig.~\ref{Fig:rmddiffnentropy}(c), the effective Hamiltonian (triangles) reproduces the RMD dynamics (circles) accurately at early times. Note, the existence of $H_{\mathrm{eff}}$ is indeed non-trivial. Conventionally, for generic many-body systems in the high-frequency limit, the effective Hamiltonian can be constructed by performing a perturbation expansion in $\omega^{-1}$. Here, our finding goes beyond this well-established framework and shows $H_{\mathrm{eff}}$ also exists in critical systems and away from the high-frequency limit.

$H_{\mathrm{eff}}$ is time-independent and hence the system does not heat up.  
The deviation of $\det (\bar M_{\eta})$ away from $1$ indicates that the orbit governed by $\bar M_{\eta}$ slowly spirals out
at a constant expansion rate that scales as $K^{{2\eta}+2}$. Further, we also note that the leading order contribution to the eigenvalues of $D_{\eta}$ scale as $K^{{\eta}+1}$, which induces a stochastic drift and the corresponding diffusion constant also scales as $K^{{2\eta}+2}$, see details
in Ref.~\cite{2024PhRvB.109f4305Y}.
Its inverse determines the prethermal lifetime scaling Eq.~\eqref{Eq:fixedscaling}, in accordance with the numerical results, Fig.~\ref{Fig:rmddiffnentropy}(b).

\subsection{Preimages}

\begin{figure}[t]
\centering
\includegraphics[width=1\linewidth]{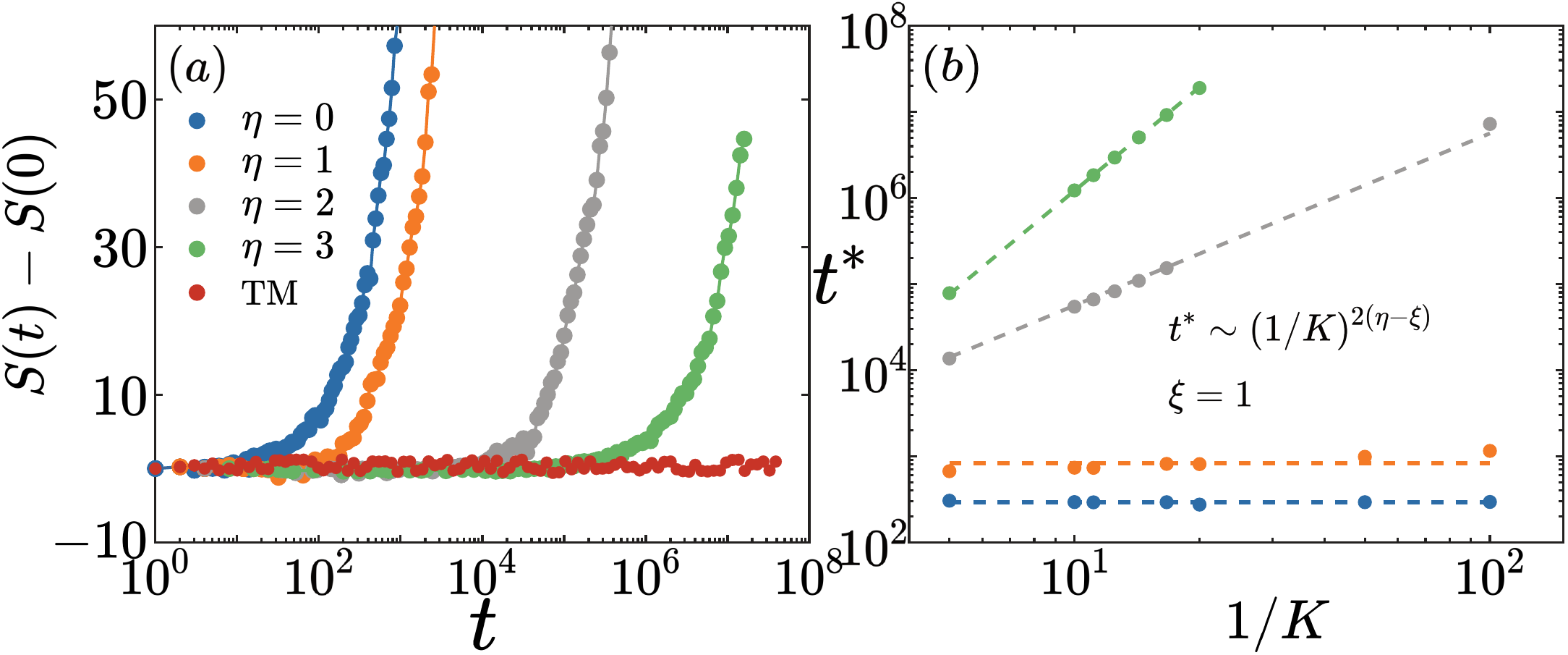}
\caption{(a) Growth of entanglement entropy for $\eta$-RMD and TM drive with $K=0.08$ for initial conditions near the first-order preimage. 
As $\eta$ increases, the prethermal lifetime prolongs.
(b) The scaling relation between the prethermal lifetime $t^*$ and $K^{-1}$ is shown on a log-log scale. Dashed lines with slopes  $2(\eta-\xi)$ are plotted as a guide. When $\eta \le \xi$, the system heats up at a constant rate. For $\eta > \xi$, the slope becomes positive and increases for larger $\eta$. We set \( T_0 / L = 2/3, \xi=1 \) in numerical simulations. 
}
\label{Fig:y=2scaling}
\end{figure}

We now extend our analysis to driving parameters that correspond to the $\xi-$order preimages with an integer $\xi \ge 1$. One possible way to parametrize the vicinity of the first-order preimage is using $T_1/L = {(-K + 2\cos(\pi T_0/L))\csc(\pi T_0/L)}/{2\pi}$, and the corresponding $(\mathrm{tr}(\mathcal{M}_1)^2, \mathrm{tr}(\mathcal{M}_2))$ deviates from the first-order preimage by an amount proportional to $K^2$. We note that for certain choices of $T_0/L$, the resulting $(\mathrm{tr}(\mathcal{M}_1)^2, \mathrm{tr}(\mathcal{M}_2))$ may still lie close to the fixed point, in which case the heating dynamics have already been discussed in Sec.~\ref{sec.fixedpoint}. In this section, we instead focus on driving parameters that are away from the fixed point.

\begin{figure}[t]
\centering
\includegraphics[width=1\linewidth]{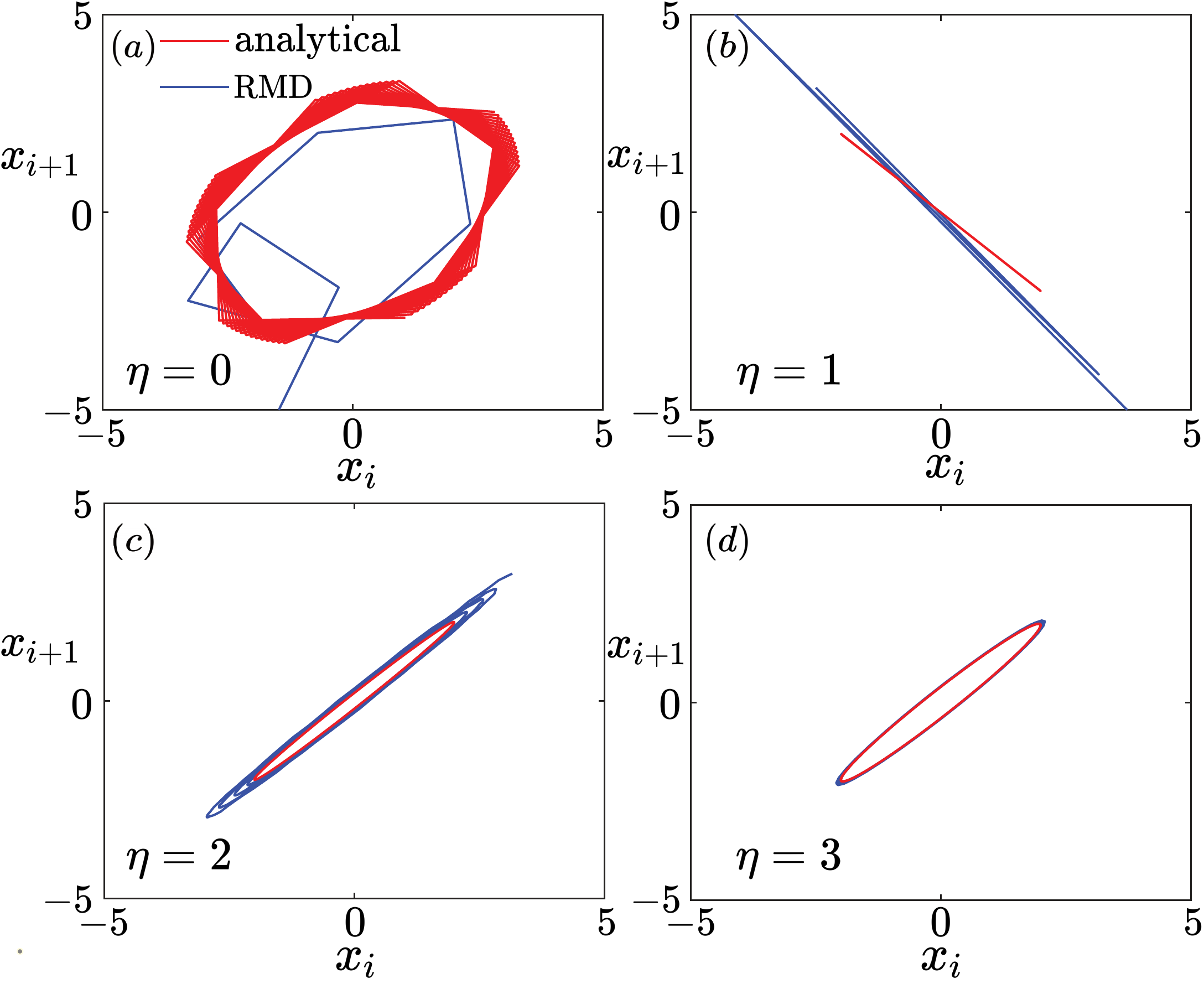}
\caption{Trace trajectory of $\eta$-RMD for driving parameters around the first-order preimages. Red curves denote the theoretical trajectory derived from the effective dynamics generated by $\bar{M}_{\eta}'$, and the blue curves denote the RMD dynamics. 
When $\eta \le \xi$, as shown in (a) and (b), the RMD trajectories notably deviate from the effective dynamics after a few steps. In contrast, when $\eta > \xi$ in panels (c) and (d), the RMD trajectories coincide with the effective trajectories. We set \( T_0 / L = 2/3, \xi=1\) for numerical simulations.}
\label{Fig:preimagetra}
\end{figure}

Fig.~\ref{Fig:y=2scaling}(a) shows the entanglement growth where we fix $\xi=1$ and change the multipolar order $\eta$. Fig.~\ref{Fig:y=2scaling}(b)
 shows the scaling of the prethermal lifetime $t^*$ versus ${K}^{-1}$. For \(\eta > 1\), the prethermal lifetime increases for larger ${K}^{-1}$ and the scaling exponent depends on the multipolar order \(\eta\), similar to the observation in Fig.~\ref{Fig:rmddiffnentropy} (b). In contrast, notably, for $\eta=0$ and 1, the system heats up at a constant rate.
We also analyze higher-order preimages and find the triply tunable prethermal lifetime \( t^* \),
\begin{align}
    t^* \sim {K} ^{-2(\eta - \xi) }, \quad \eta > \xi, \label{Eq:scalinggeneral}
\end{align}
while for $\eta\le\xi$, $t^*$ is approximately a constant.

A simple argument accounts for this behavior. For $\eta > \xi$, within a single building block ($\mathcal{M}_{\eta}$ or $\mathcal{N}_{\eta}$), the iteration relation Eq.~\eqref{Eq:tracemap2d} remains valid and maps the initial condition $(\mathrm{tr}(\mathcal{M}_1)^2, \mathrm{tr}(\mathcal{M}_2))$ to the vicinity of the fixed point. 
The subsequent heating rate analysis thus reduces to the fixed-point case discussed in Sec.~\ref{sec.fixedpoint}, and the prethermal lifetime scales as $K^{-2\eta'-2}$, where $\eta' = \eta - (\xi+1)$, the same as Eq.~\eqref{Eq:scalinggeneral}.
In contrast, when $\eta \le \xi$, a single multipolar operator is no longer sufficient to map the initial condition to the vicinity of the fixed point, but likely to a chaotic region (blue area in Fig.~\ref{Fig:preimageandentropy}).
Consequently, the system heats up rapidly regardless of the temporal correlation. Perturbatively, the average matrix $\bar{M}_{\eta}$ satisfies
\begin{align}
    \det(\bar{M}_{\eta})-1 \approx 
    \begin{cases}
        \text{constant}, & \eta \le \xi, \\
        \gamma K^{2(\eta - \xi)}, & \eta > \xi,
    \end{cases}
\end{align}
and the leading-order stochastic term $D_{\eta}$ scales as $K^{\eta - \xi}$. Therefore, for $\eta > \xi$, one expects the scaling behavior described by Eq.~\eqref{Eq:scalinggeneral}. 

This phenomenon is also visible in the evolution of the trace trajectories. As shown in Fig.~\ref{Fig:preimagetra}(a) and (b), where we use a small multipolar order $\eta=0,1$ and $\xi=1$, the RMD protocol generates the blue trajectories which quickly deviate from the effective circulating orbit (red). A notably different behavior appears for $\eta = \xi+1$: as illustrated in Fig.~\ref{Fig:preimagetra}(c), where the RMD trajectories begin to align with the effective dynamics (red), and deviations accumulate gradually at a slow pace. Further increasing $\eta$ continues to stabilize the dynamics, as shown in panel (d), where the deviation between the red and blue curves is barely visible in our numerical simulation.

\section{Heating-nonheating phase diagram in non-Hermitian systems}\label{sec:VI}
So far, we have focused on driving parameters in Region III, where the measure-zero preimages strictly forbid heating. In this section, we show that it is possible to enter the compact Region I by introducing non-unitary SU(2) deformed Hamiltonians. We design driving protocols combining both \slr\ and SU(2) deformed Hamiltonians. These realize a phase transition between a heating and a non-heating phase of non-zero measure. Remarkably, both TM and RMD protocols feature the same phase diagram, which can be explained by identifying an emergent compact group structure.

\subsection{Thue-Morse driving}
\label{sec:nonhermitian-heating}
SU(2) conformal transformations can be realized by considering the spatially deformed Hamiltonian Eq.~\eqref{Eq:Hamiltonian} with complex parameters, $\sigma^0=\cos\Gamma,\sigma^+=i \sin \Gamma,\sigma^-=0$~\footnote{This deformation is referred to as SU(2)-type, since by absorbing the imaginary unit into the Virasoro generators, one readily finds that the set \( \{L_0,\, iL_+,\, iL_-\} \) forms the generators of the SU(2) Lie algebra.
  }. We consider periodic boundary condition and $r>1$~\footnote{The above parameters do not satisfy the open boundary constraint $f_1(x)T(x)=\bar f_1(x)\bar T(x)$ at $x=0, L$, which implies no momentum flow across the boundary.}. 
This Hamiltonian turns out to be non-Hermitian, and one can analytically obtain the holomorphic and anti-holomorphic M\"obius matrices which belong to the SU(2) group (see derivations in Appendix.~\ref{part:SU(2)Hamiltonianand mobius}):
\begin{eqnarray*}
    \begin{aligned}
   U_3(T)&{=}\begin{pmatrix}
\cos(\frac{\pi T}{l}){+}i\cos\Gamma\sin\frac{\pi T}{l}   &-\sin\frac{\pi T}{l}\sin\Gamma \\
 \sin\frac{\pi T}{l}\sin\Gamma &\cos(\frac{\pi T}{l}){-}i\cos\Gamma\sin\frac{\pi T}{l}   
\end{pmatrix},\\
\tilde    U_3(T)&{=}\begin{pmatrix}
\cos(\frac{\pi T}{l}){+}i\cos\Gamma\sin\frac{\pi T}{l}   & \sin\frac{\pi T}{l}\sin\Gamma \\
- \sin\frac{\pi T}{l}\sin\Gamma &\cos(\frac{\pi T}{l}){-}i\cos\Gamma\sin\frac{\pi T}{l}   
\end{pmatrix},
\end{aligned}
\end{eqnarray*}
which differ in the sign of the off-diagonal terms.
The trace of a SU(2) matrix is constrained in $[-2,2]$.
Therefore, if the Hamiltonian parameter $\sigma^0$ alternates between two different $\Gamma$ values following the TM sequence, the trace of $\mathcal{M}_n$ always remains bounded in Region I and the Lyapunov exponent vanishes. In fact, this statement always holds regardless of the concrete driving protocol, even if it is purely random. This happens simply because the SU(2) group is compact, and hence it can never exhibit a diverging trace.

To make the system less trivial, we propose the driving protocol which involves both SU(2) and \slr\ deformed Hamiltonians, with the corresponding M\"obius matrices belonging to the non-compact \slc\ group.
Concretely, we consider the following two time evolution operators (holomorphic) 
\begin{equation}
\mathcal{M}_0{=}U_2(\lambda l)U_0(T_0) U_3(T_0),\ \mathcal{N} _0{=}U_2(\lambda l )U_0(T_1) U_3(T_1),\  \label{Eq:combinedchiral}
\end{equation}
where $\lambda$ is real,  $T_0/l=1/2+\Delta,T_1/l=1/2-\Delta$ with real $\Delta$, and
\begin{eqnarray}
    U_2(T)=\begin{pmatrix}
  \cosh(\pi\frac{T}{l})& i\sinh(\pi\frac{T}{l})\\
  -i\sinh(\pi\frac{T}{l})& \cosh(\pi\frac{T}{l})
\end{pmatrix},
\end{eqnarray} which is generated by the \slr\ spatial deformed Hamiltonian with $\sigma^0{=}\sigma^-{=}0,\sigma^+{=}1$. $\mathcal{M}_0, \mathcal{N}_0$ are now \slc, with a trace that locates them in Regions I and II.
Similarly, for the anti-holomorphic part, the M\"obius matrices are
\begin{equation}
     \tilde  {\mathcal{M}}_0{=}  U_2(\lambda l)  U_0(T_0)  \tilde U_3(T_0),
    \tilde {\mathcal{N} }_0{=}  U_2(\lambda l)  U_0(T_1)  \tilde U_3(T_1).\ \ \label{Eq:combinedantichiral}
\end{equation}
These building blocks all have a real trace and hence the classification of the invariant regions in Sec.~\ref{sec:classification} is applicable. However, for generic matrices in the \slc group, this special property does not hold. To our knowledge, a systematic classification of the 
invariant regions for a complex trace value remains an open question.

\begin{figure}[t]
\centering
\includegraphics[width=\linewidth]{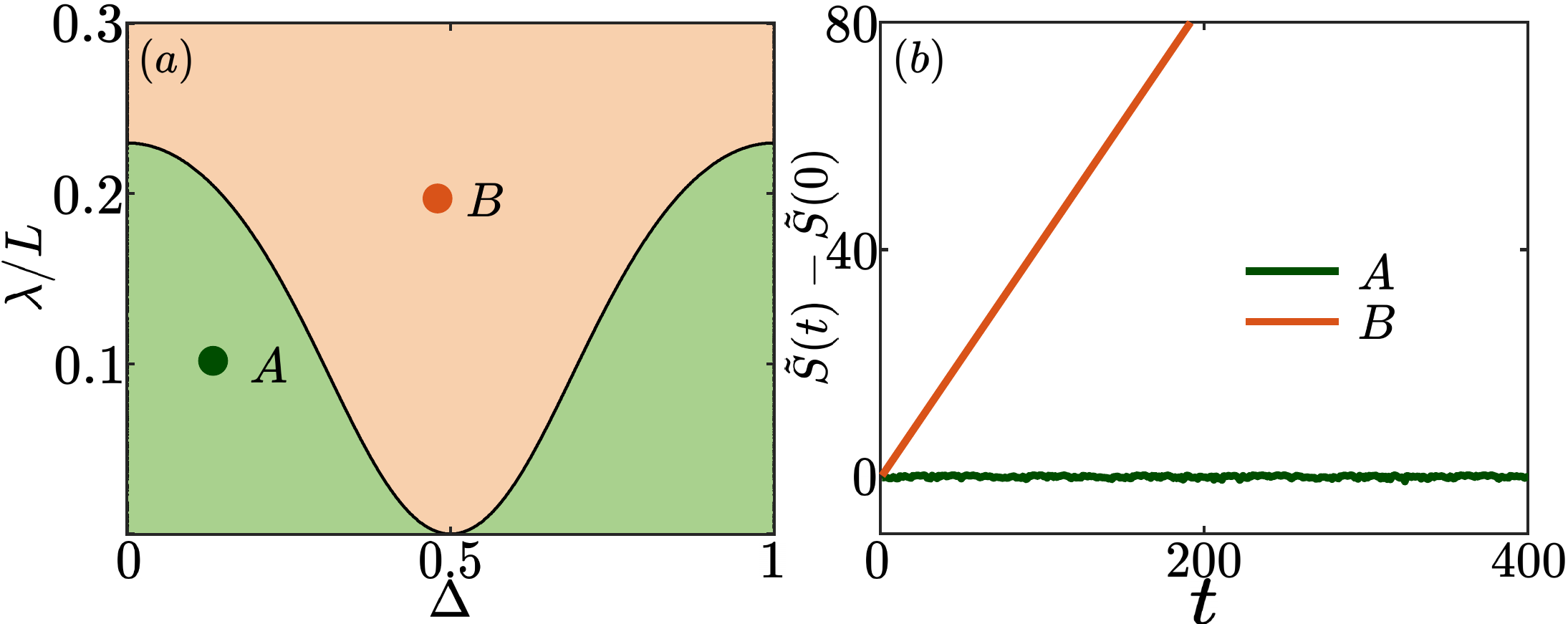}
\caption{(a) Phase diagram for the non-Hermitian driven systems, including the non-heating (green) and the heating phase (orange). (b) Growth of the pseudo entropy in different phases. For B, $(\Delta,\lambda)=(0.5,0.2)$, in the heating phase, the pseudo entropy linearly increases in time. For A, $(\Delta,\lambda)=(0.1,0.1)$, in the non-heating phase, the pseudo entropy remains close to its initial value. We set $\Gamma=\pi/2$ and take $c=1$ for numerical simulations, but the calculation works for other parameters.} 
\label{Fig:phasediagramandexample}
\end{figure}

To understand the possible phases induced by this driving protocol, we start by analyzing two limits, $\lambda=0$ and $\lambda\to\infty$. 
For $\lambda=0$, both $\mathcal{M}_0$ and $\mathcal{N} _0$ reduce to SU(2) matrices and heating does not happen. If $\lambda\rightarrow \infty$, the SU(1,1) matrix $U_2$ becomes dominant and our system is located in Region II. {Consequently, tuning $\lambda$ induces the phase transition between the non-heating (orange) and heating (green) phases in Fig.~\ref{Fig:phasediagramandexample}. The phase boundary can be analytically obtained by identifying the driving parameters corresponding to the grey line, separating Regions I and II in Fig.~\ref{Fig:TMtracemap}. This is equivalent to the condition 
    $\mathrm{tr}(\mathcal{M}_0^2\mathcal{N}_0^2)=2,$ leading to the black curve as shown in Fig.~\ref{Fig:phasediagramandexample} (a).}

For non-unitary evolution, we use the pseudo entropy as a suitable measure of the heating process~\cite{2021PhRvL.126h1601M}
\begin{eqnarray}
    \begin{aligned}
       \tilde S_m(j)&{\equiv} \frac{1}{1-m}\ln \mathrm{tr}\left [\tilde \rho_A(j)^m\right], \\
       \tilde \rho_A(j)&{=}\mathrm{tr}_{\bar A}\big [\big(\prod_{l=1}^j e^{-iH_l T_l}\big)|\psi_0\rangle\langle \psi_0|\big(\prod_{l=j}^1 e^{iH_l T_l}\big )\big],
\end{aligned}
\end{eqnarray} 
and take $m\to1$. This quantity provides a natural generalization of entanglement entropy to non-Hermitian systems within the path integral formalism. It emerges as the holographic dual to minimal area surfaces in time-dependent Euclidean spaces within the framework of AdS/CFT correspondence~\cite{2021PhRvD.103b6005N,doi2023pseudoentropy}. Subsequent studies have further used it as an order parameter to distinguish different quantum phases~\cite{2021PhRvL.126h1601M} and to separate chaotic and integrable dynamics~\cite{caputa2025thermal}. Importantly, this quantity is experimentally accessible through weak measurement techniques~\cite{nakata2021new,dressel2014colloquium}.

Using twist operators, we obtain (see details in \ref{part:realpseudoentropy})
\begin{eqnarray}
    \begin{aligned}
    \tilde S_A(j){-}\tilde S_A(0)&{=}\frac{c}{6}\ln \left( \alpha_j \gamma_j {-}\alpha_j\delta_j{-}\gamma_j\beta_j
+\beta_j\delta_j\right)\\
&+\text{antichiral},\label{Eq:psuedoSA}
\end{aligned}
\end{eqnarray}
where the central charge $c$ comes from the original undeformed unitary CFT. {We note that the pseudo entropy is generally complex in non-unitary systems. Interestingly, in our construction it remains real and positive, stemming from the fact that the M\"obius matrices, $\mathcal{M}_0, \mathcal{N}_0$, have a real trace. We illustrate the proof in Appendix. \ref{part:realpseudoentropy}}. 
We also numerically verify that the growth of the pseudo entropy exhibits the same time dependence as in the Hermitian case, $\tilde S(j)-\tilde S(0)=\frac{2c}{3}\lambda_L j$. As shown in Fig.~\ref{Fig:phasediagramandexample}(b), for Point B which locates in the orange heating phase in panel (a), the pseudo entropy increases linearly as expected. In contrast, for Point A in the green non-heating phase, the pseudo entropy stays close to its initial value.

\subsection{Random multipolar driving}
\begin{figure}[t]
\centering
\includegraphics[width=1.05\linewidth]{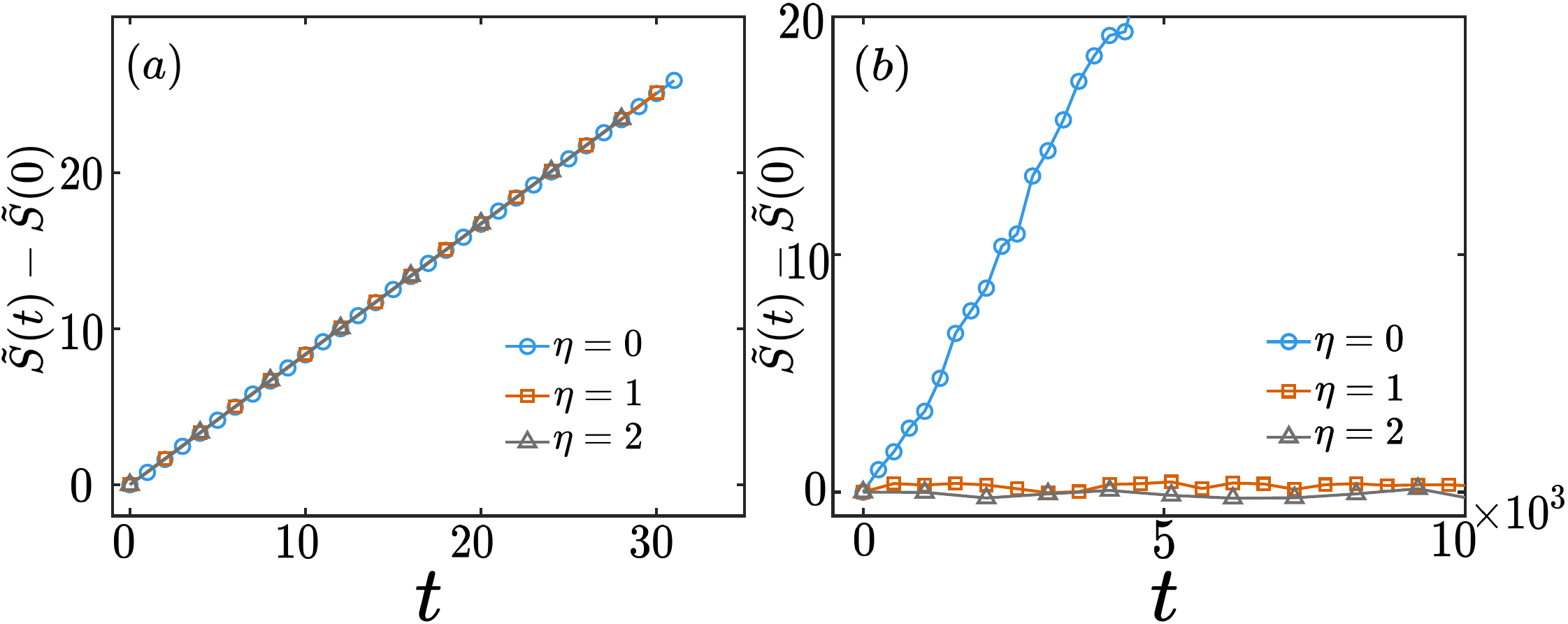}
\caption{Growth of the entanglement entropy in RMD systems. (a) For $B$ in Fig.~\ref{Fig:phasediagramandexample}, the system heats up at the same rate for different multipolar order $\eta$. (b) For $A$ in Fig.~\ref{Fig:phasediagramandexample}, $0-$RMD system heats up, while for larger multipolar order the system is non-heating. We set $\Gamma=\pi/2 
 $ and take $c=1$ for numerical simulations, but the calculation works for other parameters.}
\label{Fig:rmdheatphase}
\end{figure}

Surprisingly, the non-heating phase remains robust against temporal randomness and the phase diagram for $\eta-$RMD with $n\geq1$ is exactly the same as in Fig.~\ref{Fig:phasediagramandexample}(a). One can also verify this by calculating the pseudo entropy. As shown in Fig.~\ref{Fig:rmdheatphase}(a), Point B in Fig.~\ref{Fig:phasediagramandexample} leads to heating dynamics and crucially, 
the entanglement growth rate is independent of the multipolar order $\eta$. A notably different behavior appears for Point A: the purely random drive ($\eta=0$, blue in Fig.~\ref{Fig:rmdheatphase}(b)) results in heating; while for $\eta\geq 1$, heating is completely suppressed (red and grey) even though the driving protocol is temporally disordered.

This observation is very surprising for two reasons. First, the structure of the phase boundary in Fig.~\ref{Fig:phasediagramandexample} relies on the concomitant invariant regions of the recursive trace map, which do not exist in RMD protocols. Second, exact statements of dynamical properties in randomly driven systems, or stochastic dynamical systems, are very rare. Some exact results have been shown to exist, albeit within the framework of random matrix theory (RMT). Due to the binary multipolar construction, RMT does not naturally apply in our system.  

We account for this phenomenon by identifying a peculiar emergent compact group structure of the underlying dipolar M\"obius matrices, $\mathcal{M}_1,\mathcal{N}_1$. A similar compact structure has been originally unveiled in our previous work~\cite{mo2024non}, where the transfer matrix of non-Hermitian systems with binary disorder also exhibits this feature. Here, we sketch the core idea of the proof  and more details can be found in Appendix.~\ref{Appendix:proofsu2} and Ref.~\cite{mo2024non}.

These two matrices indeed satisfy a conjugation relation, $\mathcal{M}_1 = \sigma_z \mathcal{N}_1^* \sigma_z$, which can also be generalized to higher-order multipolar operators straightforwardly.
Using a
similarity transformation, these two matrices can be simultaneously transformed into either $\mathrm{SU}(1,1)$ or $\mathrm{SU}(2)$ matrices, depending on whether $\mathrm{tr}(\mathcal{M}_n\mathcal{N}_n)$ is greater than or less than $2$. In the entire Region I, the condition $\mathrm{tr}(\mathcal{M}_n\mathcal{N}_n) \leq 2$ is satisfied, suggesting that the product of a series of random selections of $\mathcal{M}_n$ and $\mathcal{N}_n$ is constrained in an emergent compact SU(2) manifold. Therefore, the corresponding Lyapunov exponent vanishes and heating is forbidden, in accordance with the numerical observation in Fig.~\ref{Fig:rmdheatphase}.

\section{Discussion}
\label{sec:VII}
\label{sec:randomdrive}

In this work, we investigate a family of non-periodic and structured driving protocols in CFT systems to study the heating effect in gapless quantum many-body systems. We start from a unitary \slr\ deformed CFT, where the time evolution of operators can be captured by M\"obius matrices. A recursive trace map can be obtained for the aperiodic TM protocol, which provides an efficient framework to classify the
asymptotic behavior of the entanglement growth. Properly varying the driving parameters allows us to explore different dynamical phases of matter, including the heating and non-heating phases with a fractal phase diagram~\cite{2024arXiv241018219S}, as well as the prethermal phase with a tunably slow heating rate. Using free fermion simulations, we numerically verify these results before the driving-induced high-energy excitation becomes noticeable, eventually violating the desired conformal symmetry.
We further generalize the discussion to random multipolar driving and discover a triply tunable prethermal lifetime scaling. It remains an interesting open question whether similar phenomena can be identified in conventional, locally interacting systems in the high-frequency regime.

Upon sacrificing
Hermiticity by considering SU(2) deformed CFTs, we find another non-heating phase with a non-zero measure, unreachable within the unitary \slr\  deformed CFT framework. Note, unlike the unitary CFTs being studied, it remains an interesting open question of finding a suitable lattice model that can be effectively captured by the proposed SU(2) deformed CFT model. This is of great importance in both the numerical verification and its physical realization, and we leave this for future work. 
Physically, this non-heating phase stems from 
non-Hermitian dissipation processes, which balance the energy absorption from the external drive. This naturally raises the question of how to stabilize a generic aperiodically driven interacting system via dissipation, by optimally engineering the system-bath coupling~\cite{ritter2024autonomousstabilizationfloquetstates,ma2025stabletimerondeaucrystals}. 

The emergence of a classical dynamical system $\mathcal{K}$ is essential to make the aforementioned exact predictions of the rich phase diagram. It naturally arises when analyzing the asymptotic behavior of the M\"obius transformation of operator dynamics, although the quantum many-body states follow the linear Schrodinger equation.
It is worth noting that, given a non-linear dynamical system, the classical trajectory in phase space is generally chaotic~\cite{strogatz2024nonlinear}. However, the existence of the invariant regions and the fixed point, together with the concomitant preimages, breaks the ergodicity. It further underpins the stability of the gapless quantum system, even when non-periodic time-dependence is present. Considering the everlasting studies of chaos in classical dynamical systems and the lack thereof~\cite{Russomanno2023Spatiotemporally,yan2025bifurcations}, we envision fruitful future discoveries of novel non-equilibrium quantum dynamics, if a classical non-linear map emerges in an expected way~\cite{pilatowsky2025critically}. 

We also highlight the broad applicability of Möbius matrices, or transfer matrices, in different contexts, such as in dual unitary and Gaussian quantum circuits~\cite{claeys2021ergodic,granet2023volume}, as well as in the study of Anderson localization~\cite{Anderson58,2008RvMP...80.1355E,mo2024non}. 
For example, it is suggested that the heating phase transitions in CFT systems can be mapped to a localization-delocalization transition, in both Hermitian~\cite{Lapierre2020Finestructure,PhysRevResearch.3.023044,PhysRevB.111.094304} and non-Hermitian systems~\cite{mo2024non}. {It is also intriguing to further build up the connection between prethermal phenomena and localization properties in disordered systems. }Furthermore,
generalizing the discussions to higher dimensions, both within the context of CFT~\cite{das2024exactly} or localization physics~\cite{zhou2025fundamentallocalizationphasesquasiperiodic}, is always worth pursuing. 

Finally, we note that various experimental platforms are already capable of simulating quantum many-body systems at the critical point, such as the trapped ion~\cite{joshi2023exploring} and Rydberg atom platforms~\cite{fang2024probing}. Our findings will provide enlightening insights into stabilizing those systems especially when non-periodic driving is present.

\section{Acknowledgments.---}
We thank Xueda Wen, Meng-Yang Zhang, Yingfei Gu, Jin Yan, Zhenyu Xiao, Zihan Zhou, Bingxin Lao and Bastien Lapierre for stimulating discussions. This work is supported by Quantum Science and Technology-National Science and Technology Major Project
(No. 2024ZD0301800) and by the National Natural Science
Foundation of China (Grant No. 12474214), and by “The Fundamental Research Funds for the Central Universities, Peking University”, and by ``High-performance Computing Platform of Peking University", and by the Deutsche Forschungsgemeinschaft under the cluster of excellence ct.qmat (EXC 2147, project-id 390858490) and FOR 5522 (project-id 499180199).

\appendix
\section{Free fermion numerical verification of the \slr\ CFT}\label{Appendix:free}
To numerically verify the CFT predictions, we use free-fermion simulation in a 1D lattice with the open boundary condition and $r=1$. The uniform Hamiltonian is written as 
\begin{align}
    H_0=\frac{1}{2}\sum_{j=1}^{L-1}c_j^{\dagger}c_{j+1}+\text{h.c.}
\end{align} A general \slr\ deformed Hamiltonian takes the following form 
\begin{align}
    H=\frac{1}{2}\sum_{j=1}^{L-1}f(j)c_j^{\dagger}c_{j+1}+\text{h.c.},
\end{align}
where the deformation function reads $f(j)=\sigma^0+\sigma^+\cos\frac{2\pi j}{L}+\sigma^-\sin\frac{2\pi j}{L}$ with the parameters determined by the specific driving protocols.

As shown in Fig.~\ref{Fig:preimageandentropy}(d) in the main text, we compare the growth of entanglement entropy between the CFT predictions and the free fermion numerical simulation, using three sets of different driving parameters. We observe excellent agreement for Point \( A \) and Point \( F \), while noticeable deviations appear at late times for the dynamics of Point \( D \). 
These deviations arise due to the driving-induced high-energy excitations in the lattice model, which break the exact conformal symmetry, resulting in the deviation from CFT predictions at long times.

\section{Brief Review of Conformal Field Theory}\label{Appendix:intro to CFT}

Phase transitions are a universal feature of quantum many-body systems. When a system is tuned to a continuous (second-order) critical point, the correlation length diverges and no intrinsic length scale remains. Canonical lattice examples include the transverse field Ising model, XXZ chain~\cite{lapierre2020emergent},  and free fermions at half filling~\cite{wen2018quantum}. In all such cases, scale invariance at criticality implies that the infrared (low-energy) behavior is governed by a conformal field theory; equivalently, the RG flow terminates at a conformal fixed point (see Ref.~\cite{francesco2012conformal}).

In two dimensions, the infinitesimal generators for conformal symmetry are the following holomorphic and anti-holomorphic vector fields
\begin{align}
l_n = -z^{n+1}\partial_z, \bar l_n = -\bar z^{n+1}\partial_{\bar z},n\in \mathbb{Z},
\end{align}
which satisfy the Witt algebra
\begin{align}
    [l_m,l_n]=(m-n)l_{m+n},\quad  [\bar l_m,\bar l_n]=(m-n)\bar l_{m+n},
\end{align}
and $[ l_m,\bar l_n]=0$.

Among them, $l_0,l_1,l_{-1}$ and $\bar l_0,\bar l_1,\bar l_{-1}$ generate the global conformal transformations(our focus in the main text). Specifically,  $l_{-1}$ and $\bar l_{-1}$ generate (infinitesimally) translations, $l_0+\bar l_0$ and  $i(l_0-\bar l_0)$ respectively are the generators of dilation and rotation, and $l_1,\bar l_1$ generate the special conformal transformations.

The finite form of these global generators can be described by an M\"obius transformation acting on the 2d complex coordinates
\begin{align}
    z\rightarrow \frac{az+b}{cz+d},\quad \begin{pmatrix}
 a & b\\
 c &d
\end{pmatrix}\in \text{PSL}(2,\mathbb{C}).
\end{align}
 Similarly,  we have the same transformation for the anti-holomorphic part $\bar z$. For example, the transformation for the translation $z\rightarrow z+b$ is 
 \begin{align}
     \begin{pmatrix}
 1 & b\\
 0 &1
\end{pmatrix}.
 \end{align}

Upon radial quantization, these classical generators are promoted to quantum modes $L_n,\bar L_n$ of the stress tensor and acquire a central extension, giving the Virasoro algebra, 
\begin{align}
[ L_n, L_m ] &= (n-m)\,L_{n+m}
 + \frac{c}{12}\,(n^3-n)\delta_{n+m,0},\\
[ \bar L_n,\bar L_m ] &= (n-m)\bar L_{n+m}
 + \frac{\bar c}{12}\,(n^3-n)\delta_{n+m,0},\\
[ L_n, \bar L_m ] &= 0. 
\end{align}

On a cylinder of circumference $L$ (with velocity set to $v=1$), the Hamiltonian is
\begin{align}
    H=\frac{2\pi}{L}\Big(L_0+\bar L_0-\frac{c+\bar c}{24}\Big),
\end{align}
which reduces to $\frac{2\pi}{L}(L_0+\bar L_0-\frac{c}{12})$ when $c=\bar c$. Equivalently,  one may write
\begin{align}
  H &=\int dx\mathcal{E}(x) ,
\end{align}
where the energy density is
\begin{align}
    &\mathcal{E}(x)=\frac{1}{2\pi }(T(x)+\bar T(x)),
\end{align}
where \(T\) and \(\bar T\) are the holomorphic and anti-holomorphic components of the stress–energy tensor on the cylinder.  For further details, see Refs.~\cite{ginsparg1988applied,francesco2012conformal}.

 \section{Asymptotic Growth of Entanglement Entropy}\label{part:proofeescaling}
In this section, we prove that, as $n\to\infty$, the entanglement entropy grows linearly with a rate set by the Lyapunov exponent:
\[
S_A(n)-S_A(0)\;\sim\;\frac{2c}{3}\,\lambda_L\, n.
\]

\begin{proof}
    
  Denote the two singular values of the M\"obius matrix $\Pi_n$ as $\sigma_1\ge\sigma_2\ge0$. Using
\begin{align}
        ||\Pi_n||_F\equiv \sqrt{\sigma_1(\Pi_n)^2+\sigma_2(\Pi_n)^2}\\=\sigma_1(\Pi_n)\sqrt{1+(\sigma_1(\Pi_n))^{-4}},
    \end{align}
we obtain the Lyapunov exponent as a function of singular values
\begin{align}
    \lambda_L&=\lim_{n\rightarrow\infty}\frac{1}{n}\ln (||\Pi_n||_F)\\
    &=\lim_{n\rightarrow\infty} \frac{1}{n}[\ln (\sigma_1(\Pi_n))+\frac{1}{2}\ln (1+(\sigma_1(\Pi_n))^{-4})]\\
    & = \lim_{n\rightarrow\infty} \frac{1}{n}\ln (\sigma_1(\Pi_n)),
\end{align}
where the second term of order $o(1/n)$ vanishes. 

For $\mathrm{SU}(1,1)$ matrices, one has the following Cartan decomposition,
\begin{align}
    \Pi_n = K_{1,n}A(r_n)K_{2,n},
\end{align}
where 
\begin{align}
    K_{1,n}&=\mathrm{diag}(e^{i\theta_1,n},e^{-i\theta_1,n})\\
    K_{2,n}&=\mathrm{diag}(e^{i\theta_2,n},e^{-i\theta_2,n})\\
    A(r_n)&=\begin{pmatrix}
 \cosh r_n & \sinh r_n \\
\sinh r_n  & \cosh r_n
\end{pmatrix}.
\end{align}
Therefore, $\Pi_n$ can be written as
\[
\begin{pmatrix}
 \alpha_n&\beta_n\\
\gamma_n  &\delta_n
\end{pmatrix}=\begin{pmatrix}
e^{i(\theta_1+\theta_2)} \cosh r & e^{i(\theta_1-\theta_2)} \sinh r\\
e^{-i(\theta_1-\theta_2)} \sinh r & e^{-i(\theta_1+\theta_2)} \cosh r
\end{pmatrix},
\]
whose singular value grows as
\begin{align}
    \sigma(r_n)=e^{\pm r_n}.
\end{align}
If the Lyapunov exponent is larger than zero, one has
\begin{align}
    r_n = \lambda_L n+o(n).
\end{align}
The asymptotic behavior of entanglement entropy, 
\begin{align}
  S_A(n)-S_A(0)=\frac{2c}{3}(\ln |\alpha_n-\beta_n|),
\end{align}
can be determined through
\begin{align}
    |\alpha_n-\beta_n|= |\cosh r_n-e^{-2i\theta_{2,n}}\sinh r_n|.
\end{align}
Its norm square can be determined as
\begin{align}
|\alpha_n{-}\beta_n|^2&=
\cosh^{2}  r_n {+} \sinh^{2}  r_n {-} 2 \cosh  r_n \sinh  r_n \cos \big(2\theta_{2,n}\big)\\
& =\cosh(2 r_n) {-} \cos\big(2\theta_{2,n}\big) \sinh(2 r_n)\\
& = e^{2r_n}\sin^{2}\theta_{2,n}+e^{-2r_n}\cos^{2}\theta_{2,n},
\end{align}
leading to the following expression
\begin{align}
    |\alpha_n-\beta_n|=  e^{r_n}|\sin \theta_{2,n}|\sqrt{1+e^{-4r_n}\cot^2(\theta_{2,n})}.
\end{align}
Therefore, in the long-term limit, we have the following asymptotic behavior
\begin{align}
     S_A(n)-S_A(0)=\frac{2c}{3}r_n = \frac{2c}{3}\lambda_L n,
\end{align}
which establishes the relation between the entanglement and the Lyapunov exponent, so the heating rate can be inferred from the Lyapunov exponent.
 \end{proof}

 \section{Derivation of SU(2) deformed Hamiltonian and the corresponding M\"obius transformation}~\label{part:SU(2)Hamiltonianand mobius}
Here we prove that the deformed Hamiltonian 
\begin{eqnarray}
\begin{aligned}
     H&=\frac{1}{2\pi }\int dx[T'(x)+  \bar T'(x)]\\
    &=\frac{1}{2\pi }\int dx[f_r(x) T(x)+ \bar f_r(x) \bar T(x)]
    \label{Eq:Hsu21},
\end{aligned}
\end{eqnarray}
with
$f_r(x)=\cos\Gamma+i \sin \Gamma \cos\frac{2\pi r x}{L},\sigma^0=\cos\Gamma,\sigma^+=i \sin \Gamma,$
and periodic boundary conditions lead to holomorphic and anti-holomorphic SU(2) M\"obius transformation matrices
\begin{align}
    U_1(t)&=\begin{pmatrix}
\cos(\frac{\pi t}{l})+i\cos\Gamma\sin\frac{\pi t}{l}   & -\sin\frac{\pi t}{l}\sin\Gamma \\
 \sin\frac{\pi t}{l}\sin\Gamma &\cos(\frac{\pi t}{l})-i\cos\Gamma\sin\frac{\pi t}{l}   
\end{pmatrix},\\
\tilde U_1(t)&=\begin{pmatrix}
\cos(\frac{\pi t}{l})+i\cos\Gamma\sin\frac{\pi t}{l}   & \sin\frac{\pi t}{l}\sin\Gamma \\
- \sin\frac{\pi t}{l}\sin\Gamma &\cos(\frac{\pi t}{l})-i\cos\Gamma\sin\frac{\pi t}{l}   
\end{pmatrix},\label{Eq:SU(2)mobius}
\end{align}
where $l=L/r.$
\begin{proof}
We first consider the conformal mapping
\begin{align}
  & z=\exp(\frac{2\pi r}{L}\omega),\omega=\tau+ix,\\
   &\bar z=\exp(\frac{2\pi r}{L}\bar \omega),\bar \omega=\tau-ix,\tau=it,
\end{align}
which maps the points $\omega$ on the cylinder to the $z$-plane.  One can check that
\begin{align}
    dz=\frac{2\pi r}{L}zd\omega,   d\bar z=\frac{2\pi r}{L}\bar zd\bar \omega,
\end{align}
and the Schwarzian derivative
\begin{align}
 & T(\omega)=-\frac{\pi^2 c r^2}{6L^2}+\frac{4\pi^2T(z)z^2 r^2}{L^2}, \\ 
 &T(\bar \omega)=-\frac{\pi^2 c q^2}{6L^2}+\frac{4\pi^2T(\bar z)\bar z^2 r^2}{L^2}.
\end{align}
Then we employ the conformal mapping 
\begin{align}
  &  \tilde z= \frac{\cos(\Gamma/2)z+ i\sin(\Gamma/2)}{i \sin(\Gamma/2)z+\cos(\Gamma/2)},\\
 &\bar{ \tilde z}= \frac{\cos(\Gamma/2)\bar z- i\sin(\Gamma/2)}{-i \sin(\Gamma/2)\bar z+\cos(\Gamma/2)},
\end{align}
which transforms the original Hamiltonian into a simplified form involving only the $L_0$ and $\bar{L}_0$ generators.

\begin{widetext}
Concretely,  we have
\begin{align}
    H&=\frac{1}{2\pi}\int dx (\cos\Gamma+i \sin \Gamma \cos\frac{2\pi r x}{L})T(x)+\frac{1}{2\pi}\int dx (\cos\Gamma-i \sin \Gamma \cos\frac{2\pi x}{L})\bar T(x)\\\nonumber
    &=\frac{1}{2\pi i}\int d\omega (\cos\Gamma+i \sin \Gamma \cos\frac{2\pi rx}{L})T(\omega)+\frac{1}{2\pi i}\int d\bar \omega (\cos\Gamma-i \sin \Gamma \cos\frac{2\pi rx}{L})\bar T(\bar \omega)\\\nonumber
    &=\frac{1}{2\pi i}\oint \frac{Ldz}{2\pi z r}[(\cos\Gamma+i\sin\Gamma\frac{z+z^{-1}}{2})(-\frac{\pi^2 cr^2}{6L^2}+\frac{4\pi^2T(z)z^2 r^2}{L^2})+\\\nonumber
    &\,\,\quad \frac{1}{2\pi i}\oint \frac{Ld\bar z}{2\pi \bar z r}[(\cos\Gamma-i\sin\Gamma\frac{\bar z+\bar z^{-1}}{2})(-\frac{\pi^2 c r^2}{6L^2}+\frac{4\pi^2\bar T(\bar z)\bar z^2 r^2}{L^2})]\\\nonumber
    &=\frac{2\pi r}{L}\int\frac{dz}{2\pi i}[\cos\Gamma\cdot z+\frac{i}{2}\sin\Gamma(z^2+1)]T(z)+ \text{antichiral}-\frac{\pi c}{6L}\cos\Gamma\\\nonumber
    &=\frac{2\pi r}{L}[\int \frac{d\tilde z}{2\pi i}T(\tilde z)\tilde z+\text{antichiral}]-\frac{\pi c}{6L}\cos\Gamma\\\nonumber
    &=\frac{2\pi r }{L}(L_0^{(\tilde z)}+\overline{ L}_0^{(\tilde z)})-\frac{\pi c}{6L}\cos\Gamma.
\end{align}
\end{widetext}
Then we have 
\begin{align}
  e^{iHt} \mathcal{T}^{(\tilde z)}(\tilde z,\overline {\tilde z})e^{-iHt}= \lambda^{2h}\mathcal{T}^{(\tilde z)}(\lambda \tilde z,\lambda\bar{\tilde z}),\lambda=\exp(2\pi r\tau /L).
\end{align}
which is the dilation operation in the $\tilde z$-plane.

Back to the \textit{z}-plane, its effect is to shift the operator $\mathcal{T}^{(z)}(z,\bar z)$ from $z$ to $z_{\text{new}}$, which is defined as
\begin{align}
&z_{\text{new}}= \frac{(\cos(\frac{\pi rt}{L})+i\cos\Gamma\sin\frac{\pi rt}{L} ) z-\sin\frac{\pi tr}{L}\sin\Gamma }{ \sin\frac{\pi tr}{L}\sin\Gamma z+(\cos(\frac{\pi tr}{L})-i\cos\Gamma\sin\frac{\pi tr}{L}  )},\\
&\bar z_{\text{new}}= \frac{(\cos(\frac{\pi rt}{L})+i\cos\Gamma\sin\frac{\pi rt}{L} ) z+\sin\frac{\pi tr}{L}\sin\Gamma }{ -\sin\frac{\pi tr}{L}\sin\Gamma z+(\cos(\frac{\pi tr}{L})-i\cos\Gamma\sin\frac{\pi tr}{L}  )}.
\end{align}

Therefore, the time evolution under the SU(2) deformed Hamiltonian can be written as
\begin{align*}
&\langle G | e^{H\tau} \mathcal{T}_n^{(w)}(w_0, \bar{w}_0) e^{-H\tau} | G \rangle 
= \nonumber\\& \left(\frac{\partial z}{\partial w}\right)^h \left(\frac{\partial \bar{z}}{\partial \bar{w}}\right)^h 
    \left(\frac{\partial z_{\text{new}}}{\partial z}\right)^h \left(\frac{\partial \bar{z}_{\text{new}}}{\partial \bar{z}}\right)^h \left\langle \mathcal{T}_n^{(z)}(z_{\text{new}}, \bar{z}_{\text{new}}) \right\rangle.\label{Eq:timeevolution}
\end{align*}

\end{proof}

\section{Derivation of the pseudo entropy}~\label{part:realpseudoentropy}
In the following, we are going to derive a general expression for pseudo entropy defined as
\begin{align}
    \tilde S_A^{m}\equiv \frac{1}{1-m}\ln  [\mathrm{tr}_{\bar A}(\varrho^{\psi|\phi})]^m,
\end{align}
where
\begin{align}
    \varrho^{\psi|\phi}\equiv \frac{|\psi\rangle\langle \phi|}{\langle \phi|\psi\rangle}.
\end{align}
Note that when $|\phi\rangle =|\psi\rangle$, this quantity becomes the ordinary entanglement entropy. Specifically, we define
\begin{align}
&|\psi\rangle \equiv \left (\prod_{l=1}^j e^{-iH_l T_l}\right )|\psi_0\rangle,\\
& \langle \phi|\equiv \langle \psi_0|\left (\prod_{l=j}^1 e^{iH_l T_l}\right ),
\end{align}
where $|\psi_0\rangle$ is the CFT vacuum state. When $H_l$ are Hermitian, we have $(|\psi\rangle)^{\dagger}=\langle \phi|$. But in the derivation, we consider the general case.
\begin{widetext}
In terms of the twist operator, the pseudo entropy can be written as~\cite{2021PhRvD.103b6005N}, 
\begin{align}
\tilde S_m(j)-\tilde S_m(0)&=\frac{1}{1-m}\ln \frac{\langle \psi_0|\left (\prod_{l=j}^1 e^{iH_l T_l}\right )\mathcal{T}_m(x_1,0)\mathcal{T}_m(x_2,0)\left (\prod_{l=1}^j e^{-iH_l T_l}\right )|\psi_0\rangle}{\langle \psi_0|\mathcal{T}_m(x_1,0)\mathcal{T}_m(x_2,0)|\psi_0\rangle\langle \psi_0|\left (\prod_{l=j}^1 e^{iH_l T_l}\right )\left (\prod_{l=1}^j e^{-iH_l T_l}\right )|\psi_0\rangle}\\
&=\frac{1}{1-m}\ln \frac{\langle \psi_0|\left (\prod_{l=j}^1 e^{iH_l T_l}\right )\mathcal{T}_m(x_1,0)\mathcal{T}_m(x_2,0)\left (\prod_{l=1}^j e^{-iH_l T_l}\right )|\psi_0\rangle}{\langle \psi_0|\mathcal{T}_m(x_1,0)\mathcal{T}_m(x_2,0)|\psi_0\rangle}\label{Eq:pseudoentropyderive},
\end{align}
where $\mathcal{T}_m$ and $\bar  {\mathcal{T}}_m$ are primary operators with the conformal dimension $h=\bar h=\frac{c}{24}(m-\frac{1}{m})$.
\end{widetext}
We further define $\omega_i = 0+i x_i,$ $z_i=\exp(2\pi q\omega_i/L),\zeta_i=(\tilde z_i^{(j)})^{1/q}$ and
\begin{align}
&\tilde z_i^{(j)}= \frac{\alpha_j z_i+\beta_j}{\gamma_j z_i+\delta_j},    \tilde z_i^{(j)}=\frac{\alpha_j' z_i+\beta_j'}{\gamma_j' z_i+\delta_j'}\label{Eq:generaltransform},i\in\{1,2\},
\end{align}
such that the time evolution generated by a series of Hamiltonians $\{H_l,j\in [1,j]\}$
is encoded in the M\"obius transformation. If the evolution is unitary, the M\"obius transformation belongs to the SU(1,1) group and the pseudo entropy turns into the entanglement entropy. We also consider a simple choice of the subsystem $A=[(k-1/2)l,(k+\mathcal{Q}-1/2)l],k,\mathcal{Q}\in \mathbb{Z},\mathcal{Q}< q, l=L/q$, so that the length of the subsystem is an integer times of the wavelength $l$, e.g., $|x_1-x_2|=\mathcal{Q}\cdot l$.  
\begin{widetext}
Based on this notation, the numerator in Eq.~\eqref{Eq:pseudoentropyderive} can be simplified
\begin{align}
\langle \psi_0|&\left (\prod_{l=j}^1 e^{iH_l T_l}\right )\mathcal{T}_m(x_1,0)\mathcal{T}_m(x_2,0)\left (\prod_{l=1}^j e^{-iH_l T_l}\right )|\psi_0\rangle\\
   =& \langle \psi_0 |e^{iH_jT_j}...e^{iH_1T_1}     \mathcal{T}_m(x_1,0)e^{-iH_1T_1}  ...e^{-iH_nT_n} e^{iH_jT_j}...e^{iH_1T_1}\bar{\mathcal{T}}_m(x_2,0)e^{-iH_1T_1}...e^{-iH_nT_n} | \psi_0\rangle\\
=&\prod_{i=1,2} \left( \frac{\partial \zeta_i}{\partial z_i} \right)^h \prod_{i=1,2} \left( \frac{\partial  z_i}{\partial \omega_i} \right)^h 
\prod_{i=1,2} \left( \frac{\partial \bar{\zeta}_i}{\partial \bar{z}_i} \right)^{\bar{h}} 
\prod_{i=1,2} \left( \frac{\partial \bar{ z}_i}{\partial \bar{w}_i} \right)^{\bar{h}} 
\langle \mathcal{T}(\zeta_1, \overline{\zeta_1}) \mathcal{T}(\zeta_2, \overline{\zeta_2}) \rangle\\
=&\left( \frac{2\pi}{L} \right)^{2h} 
\frac{z_1^h}{\left( \gamma_j z_1 + \delta_j \right)^{2h}} 
\frac{z_2^h}{\left( \gamma_j z_2 + \delta_j \right)^{2h}} 
\left( \frac{\alpha_j z_1+\beta_j}{\gamma_j z_1+\delta_j} \right)^{- h} 
\left(  \frac{\alpha_j z_2+\beta_j}{\gamma_j z_2+\delta_j} \right)^{- h}
\label{Eq:complexeq1}
\\
&\times \left( \frac{\alpha_j z_1+\beta_j}{\gamma_j z_1+\delta_j} \right)^{\frac{1}{q} h} \left(  \frac{\alpha_j z_2+\beta_j}{\gamma_j z_2+\delta_j} \right)^{\frac{1}{q} h}\left[ \left( \frac{\alpha_j z_1+\beta_j}{\gamma_j z_1+\delta_j} \right)^{\frac{1}{q}} 
- \left( \frac{\alpha_j z_2+\beta_j}{\gamma_j z_2+\delta_j} \right)^{\frac{1}{q}} \right]^{-2h}\label{Eq:complexeq2}\\
&\times\text{antichiral}\label{Eq:complexeq3}.
\end{align}

For the chiral part,  Eq.~\eqref{Eq:complexeq1}  can be further simplified as
\begin{align}
  &  \left( \frac{2\pi}{L} \right)^{2h} 
\frac{z_1^h}{\left( \gamma_j z_1 + \delta_j \right)^{2h}} 
\frac{z_2^h}{\left( \gamma_j z_2 + \delta_j \right)^{2h}} 
\left( \frac{\alpha_j z_1+\beta_j}{\gamma_j z_1+\delta_j} \right)^{- h} 
\left(  \frac{\alpha_j z_2+\beta_j}{\gamma_j z_2+\delta_j} \right)^{- h}\\
=& \left( \frac{2\pi}{L} \right)^{2h} 
\frac{z_1^h}{\left( \gamma_j z_1 + \delta_j \right)^{2h}} 
\frac{z_2^h}{\left( \gamma_j z_2 + \delta_j \right)^{2h}} 
\left( \frac{(\alpha_j z_1+\beta_j)(\gamma_j z_1+\delta_j)}{(\gamma_j z_1+\delta_j)^2} \right)^{- h} 
\left(  \frac{(\alpha_j z_2+\beta_j)(\gamma_j z_2+\delta_j)}{(\gamma_j z_2+\delta_j)^2} \right)^{- h}\\
=& \left( \frac{2\pi}{L} \right)^{2h} 
\frac{(-1)^h}{\left( (-\alpha_j +\beta_j)(-\gamma_j +\delta_j) \right)^{h}} 
\frac{(-1)^h}{\left((-\alpha_j +\beta_j)(-\gamma_j +\delta_j)\right)^{h}} \\
=& \left( \frac{2\pi}{L} \right)^{2h} 
\frac{1}{\left( (-\alpha_j +\beta_j)(-\gamma_j +\delta_j) \right)^{2h}} \\
=& \left( \frac{2\pi}{L} \right)^{2h} 
\frac{1}{\left( \alpha_j \gamma_j -\alpha_j\delta_j-\gamma_j\beta_j+\beta_j\delta_j\right)^{2h}} .
\end{align}
\end{widetext}

\begin{widetext}

Now let us deal with Line.~\ref{Eq:complexeq1}. We note that $z_1=z_2=-1$ , but they correspond to different layers in the $q$-sheet Riemann surface: 
\begin{align}
     \left( \frac{\alpha_j z_2+\beta_j}{\gamma_j z_2+\delta_j} \right)^{\frac{1}{q} }
     &=  \left( \frac{\alpha_j z_1+\beta_j}{\gamma_j z_1+\delta_j} \right)^{\frac{1}{q} } \exp{2\pi i\frac{\mathcal{Q}l}{L}}\\
     &=  \left( \frac{\alpha_j z_1+\beta_j}{\gamma_j z_1+\delta_j} \right)^{\frac{1}{q} } \exp{2\pi i\frac{|x_1-x_2|}{L}}.
\end{align}

Then,  Eq.~\eqref{Eq:complexeq2} can be written as 
\begin{align}
 &   \left( \frac{\alpha_j z_1+\beta_j}{\gamma_j z_1+\delta_j} \right)^{\frac{1}{q} h} \left(  \frac{\alpha_j z_2+\beta_j}{\gamma_j z_2+\delta_j} \right)^{\frac{1}{q} h}\left[ \left( \frac{\alpha_j z_1+\beta_j}{\gamma_j z_1+\delta_j} \right)^{\frac{1}{q}} 
- \left( \frac{\alpha_j z_2+\beta_j}{\gamma_j z_2+\delta_j} \right)^{\frac{1}{q}} \right]^{-2h}\\
=&\left( \frac{\alpha_j z_1+\beta_j}{\gamma_j z_1+\delta_j} \right)^{\frac{1}{q}2h } \exp{[2\pi i\frac{|x_1-x_2|}{L}\cdot h]} \left( \frac{\alpha_j z_1+\beta_j}{\gamma_j z_1+\delta_j} \right)^{-\frac{1}{q}2h } \left(1-\exp{2\pi i\frac{|x_1-x_2|}{L}}\right)^{-2h}\\
=&\frac{1}{\left (-2 i\sin \frac{\pi |x_1-x_2|}{L}\right )^{2h}}
\end{align} 

In short, the chiral part in Eq.~\eqref{Eq:complexeq1} and  Eq.~\eqref{Eq:complexeq2} can be written as
\begin{align}
     \left( \frac{2\pi}{L} \right)^{2h} 
\frac{1}{\left( \alpha_j \gamma_j -\alpha_j\delta_j-\gamma_j\beta_j+\beta_j\delta_j\right)^{2h}}\frac{1}{\left (-2 i\sin \frac{\pi |x_1-x_2|}{L}\right )^{2h}}.
\end{align}

Similarly, the antichiral part in Eq.~\eqref{Eq:complexeq3} contributes as 
\begin{align}
     \left( \frac{2\pi}{L} \right)^{2h} 
\frac{1}{\left( \alpha'_j \gamma'_j -\alpha'_j\delta'_j-\gamma'_j\beta'_j+\beta'_j\delta'_j\right)^{2h}} \frac{1}{\left (2 i\sin \frac{\pi |x_1-x_2|}{L}\right )^{2h}}.
\end{align}

Therefore, the numerator in Eq.~\eqref{Eq:pseudoentropyderive} can be written as

\begin{align}
\left( \frac{2\pi}{L} \right)^{4h} 
\frac{1}{\left( \alpha_n \gamma_n -\alpha_n\delta_n-\gamma_n\beta_n+\beta_n\delta_n\right)^{2h}} 
\frac{1}{\left( \alpha'_n \gamma'_n -\alpha'_n\delta'_n-\gamma'_n\beta'_n+\beta'_n\delta'_n\right)^{2h}}\frac{1}{\left (2\sin \frac{\pi |x_1-x_2|}{L}\right )^{4h}}.
\end{align}
Similarly, the denominator in Eq.~\eqref{Eq:pseudoentropyderive} can be written as
\begin{align}
\left( \frac{2\pi}{L} \right)^{4h} \frac{1}{\left (2\sin \frac{\pi |x_1-x_2|}{L}\right )^{4h}}.
\end{align}
In summary, 
\begin{align}
     \tilde   S_A^{(m)}(j)-\tilde S_A^{(m)}(0)
    &=\frac{c(1+m)}{12m}\left [\ln \left( \alpha_j \gamma_j -\alpha_j\delta_j-\gamma_j\beta_j+\beta_j\delta_j\right)\left( \alpha'_j \gamma'_j -\alpha'_j\delta'_j-\gamma'_j\beta'_j+\beta'_j\delta'_j\right)\right].
\end{align}
Taking the limit $m\rightarrow 1$, we have
\begin{align}
       \tilde      S_A(j)-\tilde S_A(0)   &=
    \frac{c}{6}\left [\ln \left( \alpha_j \gamma_j -\alpha_j\delta_j-\gamma_j\beta_j+\beta_j\delta_j\right)\left( \alpha'_j \gamma'_j -\alpha'_j\delta'_j-\gamma'_j\beta'_j+\beta'_j\delta'_j\right)\right].
\end{align}
 \end{widetext}
\subsection{Entropy for SU(1,1) M\"obius matrix }

If the Hamiltonian is Hermitian, the M\"obius matrix for the transformation in Eq.~\eqref{Eq:generaltransform} are SU(1,1) matrices, e.g., $\begin{bmatrix}
  \alpha_j & \beta_j\\
\gamma_j  &\delta_j 
\end{bmatrix}\in\text{SU(1,1)}$ such that their matrix components have the following relation
\begin{align}
\begin{bmatrix}
  \alpha_j & \beta_j\\
\gamma_j  &\delta_j 
\end{bmatrix}\sim\begin{bmatrix}
  \alpha_j & \beta_j\\
\beta_j^*  &\alpha_j ^*
\end{bmatrix},\begin{bmatrix}
  \alpha_j' & \beta_j'\\
\gamma_j'  &\delta_j'
\end{bmatrix}\sim\begin{bmatrix}
  \alpha_j' & \beta_j'\\
\beta_j'^*  &\alpha_j '^*
\end{bmatrix}.
\end{align}
The pseudo entropy turns out to be entanglement entropy as
\begin{align}
        S_A(j)-S_A(0)=\frac{c}{3}(\ln |\alpha_j-\beta_j| +\ln |\alpha_j'-\beta_j'|).
\end{align}
If the deformation for the chiral and antichiral in the Hamiltonian are the same, i.e., $f_r(x)=g_r(x)$, the holomorphic and anti-holomorphic M\"obius matrices are the same. Hence, the entanglement entropy becomes
\begin{align}
        S_A(j)-S_A(0)=\frac{2c}{3}\ln |\alpha_j-\beta_j| .
\end{align}

\subsection{Entropy for SU(2) M\"obius matrix}
\ref{part:SU(2)Hamiltonianand mobius} shows the M\"obius transformation under SU(2) deformed Hamiltonian evolution
\begin{align}
\begin{bmatrix}
  \alpha_j & \beta_j\\
\gamma_j  &\delta_j 
\end{bmatrix}\sim  \begin{bmatrix}
  \alpha_j & \beta_j\\
-\beta_j^*  &\alpha_j ^*
\end{bmatrix},\begin{bmatrix}
  \alpha_j' & \beta_j'\\
\gamma_j'  &\delta_j'
\end{bmatrix}\sim\begin{bmatrix}
  \alpha_j & -\beta_j\\
\beta_j^*  &\alpha_j ^*
\end{bmatrix},
\end{align}
leading to 
\begin{align}
   \tilde  S_A(j)-\tilde S_A(0)=\frac{c}{3}\ln  |-\alpha\beta^*+\beta\alpha^*-|\alpha|^2+|\beta|^2|.
\end{align}

\subsection{Entropy for the designed protocols involves SU(2) and \slr\ M\"obius matrices}
In the protocols involving SU(2) and \slr\ M\"obius matrices in Eq.~\eqref{Eq:combinedchiral} in the main text, we always have the following relations between the coefficients:
\begin{align}
    \alpha_j' = \delta_j^*, \quad \beta_j' = \gamma_j^*, \quad \gamma_j' = \beta_j^*, \quad \delta_j' = \alpha_j^*.
\end{align}

\begin{proof}
To establish these relations, we first observe that the holomorphic and anti-holomorphic M\"obius matrices in the combination protocol satisfy the following symmetry:
\begin{align}
    M_0 = \sigma_x \tilde{M}_0^* \sigma_x, \quad M_1 = \sigma_x \tilde{M}_1^* \sigma_x,
\end{align}
where \( M_0, M_1 \) are the holomorphic M\"obius matrices and \( \tilde{M}_0, \tilde{M}_1 \) are the corresponding anti-holomorphic M\"obius matrices.

This symmetry extends to any analytic function or matrix product series constructed from \( M_0 \) and \( M_1 \). Specifically, for any matrix function \( f \) formed by products of \( M_0 \) and \( M_1 \), we have:
\begin{align}
    f(M_0, M_1) = \sigma_x f(\tilde{M}_0, \tilde{M}_1)^* \sigma_x.
\end{align}
By examining the matrix elements of \( f(M_0, M_1) \) and \( f(\tilde{M}_0, \tilde{M}_1) \), we deduce the following relations for their respective components:
\begin{align}
    \alpha_j' = \delta_j^*, \quad \beta_j' = \gamma_j^*, \quad \gamma_j' = \beta_j^*, \quad \delta_j' = \alpha_j^*.
\end{align}
This completes the proof.
\end{proof}
Therefore, we have
\begin{align}
    \tilde S_A(j)-\tilde S_A(0)=\frac{c}{3}\ln \left | \alpha_j \gamma_j -\alpha_j\delta_j-\gamma_j\beta_j+\beta_j\delta_j\right|,
\end{align}
which is real.  The reality originates from the real trace preserved structure of M\"obius matrices. This is somewhat surprising due to many physical quantities becoming complex and hard to interpret in the non-Hermitian systems. We further numerically check that its scaling  also follows the simple rule with the Lyapunov exponent $\lambda_L$ as $\tilde S_A(j)-\tilde S_A(0)\sim2c/3\lambda_Lj$ in the long-time limit, showing that the pseudo entropy is an effective diagnostic for capturing dynamical features in non-unitary CFT.

\section{Proof of the non-heating phase diagram for random multipolar driving}\label{Appendix:proofsu2}

We now establish the conditions for the non-heating phase under random multipolar driving. The stability of this phase stems from the properties of the M\"obius transformations $\mathcal{M}_1$ and $\mathcal{N}_1$, consequently, extends to any general combination, including random multipolar configurations. Specifically, we prove that the driven system remains in the non-heating phase against random perturbations when the following three conditions are satisfied:
\begin{enumerate}
    \item $\mathcal{M}_1,\mathcal{N}_1$ are not triangular matrices, and they satisfy $ \mathcal{M}_1=\sigma_z\mathcal{N}_1^*\sigma_z$.
    \item $\mathrm{tr}(\mathcal{M}_1)\le 2$
    \item $   \mathrm{tr(} \mathcal{M}_1\mathcal{N}_1)\le 2$.
\end{enumerate}

\begin{proof}
    Condition 1 and $2$ imply that the eigenvalues of $\mathcal{M}$ and $\mathcal{N}$ are $u+iv,u-iv$ with $u^2+v^2=1.$ Thus they can be expressed as
    \begin{align}
  &  \mathcal{M}_1=P\Lambda P^{-1},P=\begin{bmatrix}
a  & b\\
c  &d
\end{bmatrix},\Lambda=\begin{bmatrix}
u+iv  & 0\\
0  &u-iv
\end{bmatrix}
\\ &\mathcal{N}_1=\sigma_zP^*\Lambda^* (P^*)^{-1}\sigma_z.
\end{align}where each column of  $P$ is an eigenvector of $ \mathcal{M}_1$. For simplicity, we consider $P$ to be normalized and $\det(P)=1$.

By construction, $P$ diagonalizes $\mathcal{M}_1$ into $\Lambda$, which is a $\mathrm{SU}(2)$ matrix. Similarly, $P$ transforms $\mathcal{N}_1$ to
\begin{align}
    P^{-1}\mathcal{N}_1P=P^{-1}\sigma_zP^*\Lambda^* (P^*)^{-1}\sigma_zP,
\end{align}
which in general, is not a $\mathrm{SU}(2)$ matrix, and we have the explicit expression
\begin{align}
    P^{-1}\sigma_zP^*=\begin{pmatrix}
bc^* + a^*d & b^*d + bd^* \\
-a^*c - ac^* & -b^*c - ad^*
\end{pmatrix}.
\end{align}
We further normalize the equation and derive the following expression
\begin{align}
    P^{-1}\sigma_zP^*/\sqrt{\det(\sigma_z)}=\begin{pmatrix}
-i(bc^* + a^*d) & -i(b^*d + bd^*) \\
i(a^*c + ac^*) & i(b^*c + ad^*)\label{Eq:normalsimi}
\end{pmatrix}.
\end{align}Note, its diagonal elements are conjugates of each other, an important feature of a generic $\mathrm{SU}(2)$ matrix. However, its off-diagonal elements can break the $\mathrm{SU}(2)$ structure.

There are three possibilities:

Case 1: All off-diagonal elements are zero. \(P^{-1}\sigma_zP^*/\sqrt{\det(\sigma_z)}\)  reduces to the identity matrix, which is naturally a $\mathrm{SU}(2)$ matrix; 

Case 2: Only one of the off-diagonal elements is non-zero. If so, $\mathcal{M}_1$ becomes a triangular matrix and hence Condition 1 is not satisfied.  

Case 3: Both off-diagonal elements are non-zero. In this case, we need to introduce an extra matrix 
\begin{align}
   W = 
\begin{pmatrix}
1 & 0 \\
0 & \sqrt{\frac{b^*d + bd^*}{a^*c + ac^*}}
\end{pmatrix}.
\end{align}
By applying a similarity transformation $W$ to Eq.~\eqref{Eq:normalsimi}, we have
\begin{align}
U&=W  P^{-1}\sigma_zP^*    W^{-1}/\sqrt{\det(\sigma_z)}\\&=\begin{pmatrix}
-i(bc^* + a^*d) &- i\sqrt{(b^*d + bd^*)(a^*c + ac^*)} \\
i\sqrt{(b^*d + bd^*)(a^*c + ac^*)} & i(b^*c + ad^*)
\end{pmatrix},
\end{align}
which is exactly a SU(2) matrix when the real term $(b^*d + bd^*)(a^*c + ac^*)\le 0$. 

Further calculations reveal that this condition implies $\mathrm{tr}(\mathcal{M}_1\mathcal{N}_1) \leq 2$, as evidenced by the following derivation:
\begin{align}
   \mathrm{tr(} \mathcal{M}_1\mathcal{N}_1)   &=2u^2+2v^2+4(b^*d + bd^*)(a^*c + ac^*)\\
   &=2+4(b^*d + bd^*)(a^*c + ac^*)\le 2.
\end{align}   
In summary, the three requirements collectively ensure that $\mathcal{M}_1$ and $\mathcal{N}_1$ can be simultaneously transformed via a similarity transformation into $\mathrm{SU}(2)$ matrices. Furthermore, numerical verification demonstrates that Condition 2 is inherently satisfied when Condition 3 holds. Thus, Conditions 1 and 3 alone suffice to guarantee that $\mathcal{M}_1$ and $\mathcal{N}_1$ admit a simultaneous similarity transformation into $\mathrm{SU}(2)$ matrices.

\end{proof}

\end{document}